\documentclass{jpp}

\usepackage{graphicx}
\usepackage{natbib}
\usepackage{amsmath}
\usepackage{amssymb}
\usepackage{mathtools}
\usepackage{mathrsfs}
\usepackage{lscape}
\usepackage{color}  
\usepackage{rotating}
\usepackage{booktabs}
\usepackage{caption}
\usepackage{ulem}
\usepackage{psfrag}
\usepackage{pstool}
\usepackage[colorlinks,urlcolor=blue,citecolor=blue,linkcolor=blue]{hyperref}

\ifCUPmtlplainloaded \else
  \checkfont{eurm10}
  \iffontfound
    \IfFileExists{upmath.sty}
      {\typeout{^^JFound AMS Euler Roman fonts on the system,
                   using the 'upmath' package.^^J}%
       \usepackage{upmath}}
      {\typeout{^^JFound AMS Euler Roman fonts on the system, but you
                   dont seem to have the}%
       \typeout{'upmath' package installed. JPP.cls can take advantage
                 of these fonts, if you use 'upmath' package.^^J}%
      }
  \else
  \fi
\fi

\ifCUPmtlplainloaded \else
  \checkfont{msam10}
  \iffontfound
    \IfFileExists{amssymb.sty}
      {\typeout{^^JFound AMS Symbol fonts on the system, using the
                'amssymb' package.^^J}%
       \usepackage{amssymb}%
         \let\leq=\leqslant
         
      }{}
  \fi
\fi

\ifCUPmtlplainloaded \else
  \IfFileExists{amsbsy.sty}
    {\typeout{^^JFound the 'amsbsy' package on the system, using it.^^J}%
     \usepackage{amsbsy}}
    {}
\fi

\newsavebox{\astrutbox}
\sbox{\astrutbox}{\rule[-5pt]{0pt}{20pt}}

\providecommand{\av}[1]{\langle{#1}\rangle}
\providecommand{\aV}[1]{\left\langle{#1}\right\rangle}

\def\<{\langle}
\def\>{\rangle}

\def\bi{\bf}

\providecommand\sgn{\text{sgn}}

\def\be{\begin{equation}}
\def\ee{\end{equation}}
\def\bea{\begin{eqnarray}}
\def\eea{\end{eqnarray}}
\def\nn{\nonumber}
\providecommand{\ms}{\noalign{\vspace{3pt plus2pt minus1pt}}}

\newfont{\myfont}{cmmib10}

\title[Wave dispersion in pulsar plasma 3]{Wave dispersion in pulsar plasma:\\ 3. Beam-driven instabilities}

\author[M. Z. Rafat, D. B. Melrose and A. Mastrano]%
{M.\ns Z. \ns R\ls A\ls F\ls A\ls T$^{1}$ \ns
D.\ns B.\ns M\ls E\ls L\ls R\ls O\ls S\ls E$^1$%
  \thanks{Email address for correspondence: donald.melrose@sydney.edu.au},\ns \and
A.\ns  M\ls A\ls S\ls \ls T\ls R\ls A\ls N\ls O$^{1}$}
\affiliation{$^1$SIfA, School of Physics, The University of Sydney, NSW 2006, Australia
}

\pubyear{2017}
\volume{?}
\pagerange{?}
\date{?; revised ?; accepted ?. - To be entered by editorial office}
\begin{document}

\maketitle

\begin{abstract}
Beam-driven instabilities are considered in a pulsar plasma assuming that both the background plasma and the beam are relativistic J\"uttner distributions. In the rest frame of the background, the only waves that can satisfy the resonance condition are in a tiny range of slightly subluminal phase speeds. The growth rate for the kinetic (or maser) version of the weak-beam instability is much smaller than has been estimated for a relativistically streaming Gaussian distribution, and the reasons for this are discussed. The growth rate for the reactive version of the weak-beam instability is treated in a conventional way. We compare the results with exact calculations, and find that the approximate solutions are not consistent with the exact results. We conclude that, for plausible parameters, there is no reactive version of the instability. The growth rate in the pulsar frame is smaller than that in the rest frame of the background plasma by a factor $2\gamma_{\rm s}$, where $\gamma_{\rm s} = 10^2 {\rm -} 10^3$ is the Lorentz factor of the bulk motion of the background plasma, placing a further constraint on effective wave growth. Based on these results, we argue that beam-driven wave growth probably plays no role in pulsar radio emission.

\end{abstract}

\begin{PACS}
\end{PACS}

\section{Introduction}

In two accompanying papers (Rafat, Melrose and Mastrano 2018a,b, referred to here as Papers~1 and~2) we discuss aspects of the plasma physics relevant to beam-driven instabilities in a pulsar plasma, defined here (as in Paper~1) to be a relativistic, one-dimensional (1D), electron-positron plasma that is streaming outward, with streaming Lorentz factor $\gamma_{\rm s} \gg 1$, where $ \av{\cdots} $ indicated the average value, and with a relativistic spread, $\langle\gamma\rangle\gtrsim1$, in its rest frame. Beam instabilities were discussed extensively in the early pulsar literature \citep[e.g.,][]{TK72,SC73,SC75,H76b,H76a,HR76,HR78,BB77,LM79,LMS79,LP82,Letal82,APS83,ELM83,L92,Lyubarsky96,A93,A95,W94,AM98,MG99,L00,AR00}. Various different assumptions were made relating to the properties of the growing waves (Langmuir-like, Alfv\'en-like), the model for the background plasma in its rest frame (cold, nonrelativistic thermal, intrinsically relativistic) and to the form of the instability (kinetic or reactive). Beam-driven wave growth continues to be invoked in the two most widely favored pulsar radio emission mechanisms (Paper~1): as the source of wave energy in relativistic plasma emission (RPE) \citep[e.g.,][]{EH16},  and as the bunching mechanism for coherent curvature emission (CCE) \citep[e.g.,][]{MGP00,MGM09}. In Paper~1 we discuss wave dispersion in the rest frame of the plasma assuming a relativistic J\"uttner distribution \citep{J11,Synge57,WH75} and emphasize that all waves have phase speed\footnote{In Papers 1 and 2, $ z = \Re \omega/c\Re k_\parallel $ which corresponds to $ \zeta $ in this paper. This inconsistency in notation is unavoidable as $ z $ is complex here.}, $\zeta=\Re \omega/\Re k_\parallel c$, either just below (subluminal) or greater than (superluminal) unity. This severely restricts beam-driven instabilities due to the difficulty of satisfying the resonance condition between a beam and any wave in the background plasma. In particular we argue that, contrary to what has often been assumed, there are no Langmuir-like waves with non-relativistic phase speeds in a pulsar plasma. In Paper~2 we argue that the default choice for the beam distribution should be a Lorentz-transformed J\"uttner distribution, rather than the widely-favored choice of a relativistically streaming Gaussian distribution, and we find that when this choice is made it implies that the distribution function is much broader that the widely-favored form would suggest. This further constrains wave growth in several ways. It decreases the positive slope of the beam distribution below its maximum, thereby reducing the growth rate for the kinetic instability, and it increases the bandwidth of the growing waves, and it effectively precludes reactive wave growth. Our purpose in this paper is to derive the growth rates for various beam-driven instabilities, and to discuss their possible application to pulsars. We conclude that the constraints are so severe that it is implausible that beam-driven wave growth plays any role in pulsar radio emission.

We compare our results with those of \cite{ELM83} (ELM), who assumed a relativistically streaming Gaussian distribution and determined the wave properties using a relativistic form of dispersion theory \citep{LM79,LMS79}. Following ELM, we estimate the growth rates for three versions of the beam instability: the kinetic (or maser) version and two reactive (or hydrodynamic) versions, that we refer to as resonant and nonresonant \citep{GMG02}. For the kinetic instability to apply, the growth rate must be less than the bandwidth of the growing waves, and this condition fails to be satisfied by many orders of magnitude under plausible conditions in a pulsar magnetosphere for the distribution assumed by ELM, who concluded that the beam instability must be reactive. We derive the growth rates for these instabilities for a Lorentz-transformed J\"uttner distribution and compare the results with those for the relativistically streaming Gaussian distribution assumed by ELM. We argue that there is an additional constraint that imposes a more severe constraint on the kinetic instability. This relates to the requirement for the beam and background distributions to be well separated (Paper~2); if this constraint is not satisfied, Landau damping by particles in the tail of the background distribution overwhelms any tendency towards negative absorption from the beam. Well-separated beam and background distributions are also required for either reactive version to apply, and we argue that this condition is not plausibly satisfied. Our conclusions, contrary to the conclusions of ELM, are that for J\"uttner distributions, kinetic growth is possible but ineffective, and that reactive growth is not possible for plausible parameters.

We discuss the weak-beam distribution in \S\ref{sect:weak-beam}.
In \S\ref{sect:kinetic} we derive and discuss the growth rate for the kinetic weak-beam instability, and in \S\ref{sect:reactive} we discuss reactive wave growth. The relation between temporal and spatial wave growth is discussed in \S\ref{sect:LTgrowth}. The results are discussed in \S\ref{sect:discussion} and concluded in \S\ref{sect:conclusions}.

\section{Weak-beam model}
\label{sect:weak-beam}

In this section we discuss the Lorentz-transformed J\"uttner model for a weak beam, and compare it with a relativistically streaming Gaussian model. We then consider dispersion equation for parallel propagating waves for a weak-beam model.

\subsection{Lorentz transformed J\"uttner distribution}

A plasma distribution consisting of a background plasma and a beam may be written as $ g(u) = g_0(u) + g_1(u) $, where $ g_0(u) $ describes the background plasma and $ g_1(u) $ describes the beam. Assuming that these are, respectively, a J\"uttner distribution at rest and a J\"uttner distribution streaming with speed $\beta_{\rm b}$ gives (Paper~2)
\begin{equation}
    g_0(u)
    = n_0\frac{e^{-\rho_0\gamma}}{2K_1(\rho_0)},\qquad
    g_1(u)
    = \frac{n_1}{\gamma_{\rm b}}\frac{e^{-\rho_1\gamma_{\rm b}\gamma(1-\beta\beta_{\rm b})}}{2K_1(\rho_1)},
\label{guwb}
\end{equation}
where $ n_0 $ and $ n_1 $ are the number densities of the background and the beam in the rest frame of the background. The distributions are normalized so that
\begin{equation}
    n_0 = \int du\, g_0(u),\qquad
    n_1 = \int du\, g_1(u).
\end{equation}
A weak beam is defined (Paper~2) as one where the number density of the beam in its rest frame, $ n_1/\gamma_{\rm b} $, is much smaller than the number density of the background in its rest frame, giving
\begin{equation}
    \varepsilon_n = \frac{n_1}{\gamma_{\rm b} n_0} \ll 1.
\end{equation}
We write the combined distribution $ g(u) $ and its derivative as
\begin{equation}\label{eq:gdgdbeta}
\begin{split}
    g(u) 
        & = \frac{n_0}{2K_1(\rho_0)}\left[e^{-\rho_0\gamma} + \varepsilon_n\varepsilon_Ke^{-\rho_1\gamma_{\rm b}\gamma(1-\beta_{\rm b}\beta)}\right],\\
    \frac{dg(u)}{d\beta}
        & = \frac{-n_0\rho_0\gamma^3}{2K_1(\rho_0)}\left[\beta e^{-\rho_0\gamma} + \varepsilon_n\varepsilon_\rho\varepsilon_K\gamma_{\rm b}(\beta - \beta_{\rm b})e^{-\rho_1\gamma_{\rm b}\gamma(1-\beta_{\rm b}\beta)}\right],
\end{split}
\end{equation}
where we define
\begin{equation}
    \varepsilon_\rho = \rho_1/\rho_0,\quad
    \varepsilon_K = K_1(\rho_0)/K_1(\rho_1),
\end{equation}
with $ \varepsilon_K \approx \varepsilon_\rho $ for $ \{\rho_0, \rho_1\} \ll 1 $ since $ 1/K_1(x) = x + \mathcal{O}(x^3) $ for $ 0<x\ll 1 $. 

As noted in Paper~2, in the highly relativistic case, $\gamma_{\rm b}\gg \{1/\rho_0,1/\rho_1\} \gg1$, one may approximate~\eqref{guwb} by
\begin{equation}
    g_0(u)
        = \frac{n_0\rho_0}{2}e^{-\rho_0\gamma},
    \quad
    g_1(u)
        = \frac{n_1\rho_1}{2\gamma_{\rm b}}\exp\left[-\frac{\rho_1(\gamma-\gamma_{\rm b})^2}{2\gamma_{\rm b}\gamma}\right].
\label{guwb1}
\end{equation}

\subsubsection{Average quantities}
In the rest frame of the background, averages of a quantity $Q$ over the background and over the beam are
\begin{equation}
    n_0\av{Q}_0 = \int du\, Q\, g_0(u),\qquad
    n_1\av{Q}_1 = \int du\, Q\, g_1(u),
\end{equation}
respectively. The average of $ Q $ over the combined distributions is
   $\av{Q} = (n_0\av{Q}_0 + n_1\av{Q}_1)/(n_0 + n_1)$.
The relation (Paper~2) $ n''_\alpha\aV{Q/\gamma''}''_\alpha = n_\alpha\aV{{Q}/{\gamma}}_\alpha $ between the averages of $ Q $ in two frames ${\cal K}$ (rest frame of the background) and ${\cal K}''$ (rest frame of the beam) may be used to relate $n_1\av{Q}_1$ to $n_0\av{Q}_0$. We denote parameters in the rest frame of the beam with double primes.

\subsubsection{Comparison with Gaussian model}

A widely favored choice for a relativistically streaming distribution is a Gaussian. Assuming a non-streaming Gaussian model for the background and a relativistically streaming Gaussian model for the beam gives
\begin{equation}
g_{\rm G0}(u) = \frac{n_0}{\sqrt{\pi}u_{\rm T0}}\exp\left[-\frac{u^2}{u_{\rm T0}^2}\right],
\qquad
g_{\rm G1}(u)= \frac{n_1}{\sqrt{\pi}\,u_{\rm Tb}}\exp\left[-\frac{(u-u_{\rm b})^2}{u_{\rm Tb}^2}\right],
\label{Gaussian}
\end{equation} 
in place of $ g_0(u) $ and $g_1(u)$. The derivative of the total distribution, $g_{\rm G}(u) = g_{\rm G0}(u) + g_{\rm G1}(u) $, is
\begin{equation}
\frac{{\rm d}g_{\rm G}(u)}{{\rm d}\beta}
    =-\frac{2n_0\gamma^3}{\sqrt{\pi}\,u_{\rm T0}^3}
    \left\{u\exp\left[-\frac{u^2}{u_{\rm T0}^2}\right] + \varepsilon_n\varepsilon_{T}^3\gamma_{\rm b}(u-u_{\rm b})\exp\left[-\frac{(u-u_{\rm b})^2}{u_{\rm Tb}^2}\right]\right\},
\label{dGaussian}
\end{equation} 
with $ \varepsilon_T = u_{\rm T0}/u_{\rm Tb} $ and $ \varepsilon_n = n_1/\gamma_{\rm b}n_0$. 

We emphasize that $g_{\rm G1}(u)$ is not the same as the distribution obtained by first setting $u_{\rm b}=0$ and then making the Lorentz transformation to the frame in which the distribution is streaming with speed $\beta_{\rm b}$. The distribution obtained by doing so is similar to the Lorentz-transformed J\"uttner distribution, both of which are much broader than $g_{\rm G1}(u)$ in the highly relativistic case.

\subsection{Dispersion equation for parallel propagation}

Following Paper~1, the dispersion equation for parallel propagation for a weak beam model may be written as
\begin{equation}
    K_{33} = 1 - \frac{\omega_{\rm p0}^2}{\omega^2}z^2W_0(z) - \frac{\omega_{\rm p1}^2}{\omega^2}z^2W_1(z),
\label{wb1}
\end{equation}
where subscripts $\alpha=0,1$ refer to the background and beam, respectively, and $ \omega_{\rm p\alpha}^2 = n_\alpha e^2/m\varepsilon_0 $ is the plasma frequency. The relativistic plasma dispersion function (RPDF) $ W_\alpha(z) $ is given by
\begin{equation}\label{eq:Wdef}
    W_\alpha(z) 
        = \frac{1}{n_\alpha}\int_{-\infty}^\infty du\, \frac{dg_\alpha(u)/du}{\beta - z}
        = \frac{1}{n_\alpha}\int_{-1}^1 d\beta\, \frac{dg_\alpha(u)/d\beta}{\beta - z},
\end{equation}
where $ z = \omega/ck_\parallel $ is the phase velocity and $ g_\alpha(u) $ are given by~\eqref{guwb}. For complex frequency $ \omega = \omega_r + i\omega_i $ and wave-number $ k_\parallel = k_r + ik_i $, we have $ z = z_r + i z_i $ with
\begin{equation}\label{eq:zrzi}
    z_r = \frac{\omega_r k_r + \omega_i k_i}{c|k_\parallel|^2},\quad
    z_i = \frac{\omega_i k_r - \omega_r k_i}{c|k_\parallel|^2}.
\end{equation}
Landau prescription implies that the integral path in the complex $ \beta $ plane runs below the poles. Thus, for $ -1 \leq z_r \leq 1 $, we have
\begin{equation}\label{eq:Wcases}
    W_\alpha(z) = \frac{1}{n_\alpha}
    \begin{dcases}
        \int_{-1}^1 d\beta\, \frac{dg_\alpha(u)/d\beta}{\beta - z}, & \quad {\rm for} \quad z_i > 0,\\
        \wp\int_{-1}^{1}{\rm d}\beta\, \frac{dg_\alpha(u)/d\beta}{\beta - z_r} + i\pi\sgn{(z_r)} \left.\frac{dg(u)}{d\beta}\right|_{\beta = z_r}, & \quad {\rm for} \quad z_i = 0,\\
        \int_{-1}^1 d\beta\, \frac{dg_\alpha(u)/d\beta}{\beta - z} + i2\pi\sgn{(z_r)} \left.\frac{dg_\alpha(u)}{d\beta}\right|_{\beta = z}, & \quad {\rm for} \quad z_i < 0,
    \end{dcases}
\end{equation}
where $ \wp $ denotes Cauchy Principal Value. Integration by parts implies that
\begin{equation}\label{eq:Wints}
\begin{split}
    \wp\int_{-1}^{1}{\rm d}\beta\, \frac{dg_\alpha(u)/d\beta}{\beta - z_r}
        & = -\left.2\gamma^2 g_\alpha(u)\right|_{\beta = z_r} - \wp\int_{-1}^{1}{\rm d}\beta\, \frac{\left. g_\alpha(u)\right|_{\beta = z_r} - g_\alpha(u)}{(\beta - z_r)^2},\\
    \int_{-1}^1 d\beta\, \frac{dg_\alpha(u)/d\beta}{\beta - z}
        & = \int_{-1}^1 d\beta\, \frac{g_\alpha(u)}{(\beta - z)^2},
\end{split}
\end{equation}
where the first relation is shown in Paper~1. We express~\eqref{wb1} as
\begin{equation}\label{eq:omegawb}
    K_{33} = 1 - \frac{\omega_{\rm p0}^2}{\omega^2}z^2W(z),\quad
    W(z) = W_0(z) + \varepsilon_n\gamma_{\rm b} W_1(z),
\end{equation}
where we use $ \omega_{\rm p1}^2 = \varepsilon_n\gamma_{\rm b}\omega_{\rm p0}^2 $ as $ n_1 = \varepsilon_n \gamma_{\rm b} n_0 $, and $ W(z) $ is calculated as above with $ n_\alpha \to n_0 $ and $ g_\alpha(u) $ replaced by the combined distribution $ g(u) $ as given by~\eqref{eq:gdgdbeta}. The form (\ref{eq:omegawb}) of the dispersion equation is useful in discussing kinetic instability. 

The RPDF associated with the beam may be evaluated either in the rest frame of the background, $ W_1(z)$, or in the rest frame of the beam, $W''_1(z'')$, with
\begin{equation}
    \frac{z^2 W_1(z)}{\omega^2} = \frac{1}{\gamma_{\rm b}} \frac{z''^2W''_1(z'')}{\omega''^2},\quad
    z'' = \frac{z - \beta_{\rm b}}{1 - \beta_{\rm b} z},\quad
    \omega'' = \gamma_{\rm b}\omega(1-\beta_{\rm b}/z),
\end{equation}
as discussed in Paper~2. We may then write the dispersion equation~\eqref{wb1} as
\begin{equation}\label{eq:K33r0}
    K_{33} 
        = 1 - \omega_{\rm p0}^2\frac{z^2W_0(z)}{\omega^2} - \varepsilon_n \omega_{\rm p0}^2\frac{z''^2 W''_1(z'')}{\gamma_{\rm b}^2(\omega - \beta_{\rm b}ck_\parallel)^2}.
\end{equation}
This form is convenient in discussing a reactive instability. 

\section{Kinetic form of weak-beam instability}
\label{sect:kinetic} 

The growth rate for the kinetic version of the weak-beam instability may be derived using two different methods, that are known to lead to equivalent results. The first method involves associating the absorption to the imaginary component of the frequency which is obtained from the dispersion equation. The second involves deriving the absorption coefficient using a semi-classical theory, using detailed balance to relate absorption to emission, and identifying wave growth as negative absorption. We develop the former method for a pulsar plasma, extending results derived in Paper~1.

\subsection{Absorption coefficient from dispersion theory}

The dissipative part of the response is described by the imaginary part of the dielectric constant. Landau damping or wave growth in a kinetic instability may be described by $\Im\omega \equiv \omega_i \ne 0 $ and  $\Im k_\parallel \equiv k_i \ne 0 $, such that wave energy varies as $\exp(2\omega_i t - 2 k_i s)$, where $s$ denotes distance along the magnetic field (which is also the ray path in the case considered here). Whether damping is temporal or spatial depends on the boundary conditions. If the waves are initially uniformly distributed then purely temporal damping ($ \omega_i \ne 0 $, $ k_i = 0 $) applies, and if the waves are generated by a constant point source then purely spatial damping ($ \omega_i = 0 $, $ k_i \ne 0 $) applies away from that point. The absorption coefficient in the form $ \Gamma_L = -2(\omega_i - c\beta_{g}k_i) $, where $ \beta_{\rm g} $ is the group speed of the wave, describes damping independent of the boundary conditions. The energy of the wave then varies as $ \exp(- t\Gamma_L) $ so that negative absorption, $ \Gamma_L < 0 $, implies that the energy of the wave increases while positive absorption, $ \Gamma_L > 0 $, implies wave damping.

The dispersion equation~\eqref{eq:omegawb} with $ K_{33}(\omega, k_\parallel) = \Re K_{33}(\omega, k_\parallel) + i\Im K_{33}(\omega, k_\parallel) $, $ \omega = \omega_r + i\omega_i $ and $ k_\parallel = k_r + ik_i $ is solved using a perturbation approach. One expands $ K_{33}(\omega, k_\parallel) $ about $ (\omega, k_\parallel) = (\omega_r, k_r)$ to first order in $ \omega_i $ and $ k_i $, assuming $ \{|\omega_i/\omega_r|, |k_i/k_r|\} \ll 1 $, i.e. weak damping/growth, which gives
\begin{equation}\label{eq:K33pert}
\begin{split}
    K_{33}(\omega, k_\parallel) 
        & \approx \Re K_{33}(\omega_r, k_r) + i\Im K_{33}(\omega_r, k_r)\\
        & + i\left.\left(\omega_i \frac{\partial \Re K_{33}(\omega, k_\parallel)}{\partial \omega_r} + k_i \frac{\partial \Re K_{33}(\omega, k_\parallel)}{\partial k_r}\right)\right|_{\omega = \omega_r, k_\parallel = k_r}.
\end{split}
\end{equation}
Equating the real part to zero, $ \Re K_{33}(\omega, k_\parallel) = 0 $, gives
\begin{equation}\label{eq:pertRe}
    \omega_r^2 = \omega^2_L(\zeta) = \omega^2_{\rm p0} \zeta^2 \Re W(\zeta),\quad
    c^2 k_r^2 = \omega^2_L(\zeta)/\zeta^2 = \omega^2_{\rm p0} \Re W(\zeta),
\end{equation}
where we denote\footnote{The phase $ z $ in Papers 1 and 2 corresponds to $ \zeta $ in this paper.} $ \zeta = \left.z\right|_{\omega = \omega_r, k_\parallel = k_r} = \omega_r/ck_r $, and $ W(\zeta) $ may obtained using~\eqref{eq:Wcases} as
\begin{equation}\label{eq:Wzeta}
    \Re W(\zeta)
        = \frac{1}{n_0}\wp\int_{-1}^1d\beta\, \frac{dg(u)/d\beta}{\beta - \zeta},\quad
    \Im W(\zeta)
        = \frac{\pi}{n_0}\sgn{(\zeta)}\left.\frac{dg(u)}{d\beta}\right|_{\beta = \zeta}.
\end{equation}
The solution of the imaginary part, $ \Im K_{33}(\omega, k_\parallel) = 0 $, gives
\begin{equation}\label{eq:pertIm}
    \omega_i - c\beta_g k_i = -\omega_L(\zeta) R_L(\zeta)\Im K_{33}(\omega_r, k_r),
\end{equation}
where the group speed $ c\beta_g = \partial\omega/\partial k_\parallel $ is
\begin{equation}\label{eq:pertBetaG}
    c\beta_{g}
    = -\left.\frac{\partial \Re K_{33}(\omega,k_\parallel)/\partial k_r}{\partial \Re K_{33}(\omega,k_\parallel)/\partial \omega_r}\right|_{\omega = \omega_r, k_\parallel = k_r},
\end{equation}
and the ratio of electric to total energy is
\begin{equation}\label{eq:pertRL}
    R_{L}(\zeta) 
    = \left.\frac{1}{\omega\partial \Re K_{33}(\omega, k_\parallel)/\partial \omega_r}\right|_{\omega = \omega_r, k_\parallel = k_r}.
\end{equation}
The explicit expressions for $ \beta_g(\zeta) $ and $ R_L(\zeta) $ are obtained as
\begin{equation}
    \beta_g(\zeta) = \frac{d\left[\zeta^2\Re W(\zeta)\right]/d\zeta}{\zeta d\Re W(\zeta)/d\zeta},
    \quad\text{and}\quad
    R_L(\zeta) = -\frac{\Re W(\zeta)}{\zeta d\Re W(\zeta)/d\zeta},
\end{equation}
with $ \beta_g(\zeta) = \zeta[1 - 2R_L(\zeta)] $. The absorption coefficient then follows from~\eqref{eq:pertIm},
\begin{equation}
    \Gamma_L(\zeta) = -2(\omega_i - c\beta_{g} k_i) = 2\omega_L(\zeta)R_L(\zeta)\Im K_{33}(\omega_r, k_r).
\label{eq:adt2b}
\end{equation}

The fractional absorption coefficient $ \overline{\Gamma}_L(\zeta) = \Gamma_L(\zeta)/\omega_L(\zeta) $ for the distribution (\ref{eq:gdgdbeta}) may be written as
\begin{equation}\label{eq:fracAbsDisp}
\begin{split}
    \overline{\Gamma}_L(\zeta) 
        & = \frac{-\sgn(k_r)}{\zeta d\Re W(\zeta)/d\zeta}\frac{\pi\rho_0\gamma_\phi^3}{K_1(\rho_0)}\left[\zeta e^{-\rho_0\gamma_\phi} + \varepsilon_n\varepsilon_\rho\varepsilon_K \gamma_{\rm b}(\zeta - \beta_{\rm b})e^{-\rho_1\gamma_{\rm b}\gamma_\phi(1 - \zeta\beta_{\rm b})}\right],
\end{split}
\end{equation}
with $ \gamma_\phi = \left.\gamma\right|_{\beta = \zeta} $ and
\begin{equation}
    \frac{d \Re W(\zeta)}{d\zeta}
        = \frac{1}{n_0}\wp\int_{-1}^1d\beta\, \frac{dg(u)/d\beta}{(\beta - \zeta)^2}.
\end{equation}
Equations~\eqref{eq:fracAbsDisp} describe damping ($ \overline{\Gamma}(\zeta) > 0 $) or growth ($ \overline{\Gamma}(\zeta) < 0 $) independent of whether it is temporal or spatial. 

The counterpart of~\eqref{eq:fracAbsDisp} for the combination of Gaussian distributions (\ref{Gaussian}) is
\begin{equation}
    \overline{\Gamma}_{GL}(\zeta)
    = \frac{-\sgn(\zeta)2\sqrt{\pi}}{d\Re W_G(\zeta)/d\zeta}\frac{ \gamma_\phi^3}{u_{\rm T0}^3}\left\{u_\phi e^{-u_\phi^2/u_{\rm T0}^2} + \varepsilon_n\varepsilon_{T}^3\gamma_{\rm b}(u_\phi-u_{\rm b})e^{-(u_\phi-u_{\rm b})^2/u_{\rm Tb}^2}\right\},
\end{equation}
where $ \overline{\Gamma}_{GL}(\zeta) = \Gamma_{GL}(\zeta)/\omega_{GL}(\zeta) $, $ u_\phi = \gamma_\phi \zeta $ and $ \omega_{GL} (\zeta) = \omega_{\rm p0}^2 \zeta^2 \Re W_{G}(\zeta) $ with $ W_G(\zeta) $ defined as $ W(\zeta) $ but with the J\"uttner distribution replaced by a Gaussian distribution.

\subsection{Separation condition}

The particles in the tail of the background distribution contribute towards positive (Landau) damping of the waves, and a negative contribution from the beam must overcome this positive contribution in order for the waves to grow. The net (background plus beam) distribution must have a well-defined minimum, with a positive slope, $ \sgn{(u)} dg(u)/du > 0 $, above the minimum (and below the maximum in the beam distribution) to drive wave growth. This is discussed in the context of Penrose criterion for instability in \S\ref{sec:Penrose} and Appendix~\ref{app:Penrose}.

The first term inside the braces in the expression~\eqref{eq:fracAbsDisp} for fractional absorption coefficient $ \overline{\Gamma}_L(\zeta) $ is the positive contribution from the background, and the second term is the contribution from the beam, which is negative for $ \gamma_\phi < \gamma_{\rm b} $ and positive for $ \gamma_\phi > \gamma_{\rm b} $.  Negative absorption, $ \overline{\Gamma}_L(\zeta) < 0 $, requires not only that the contribution of the beam be negative, $0 < \zeta < \beta_{\rm b}$ or $\gamma_\phi<\gamma_{\rm b}$, but also that the negative contribution from the beam be greater in magnitude than the positive contribution from the background. This leads to the separation condition, that the sum of the two contributions, $\propto dg(u)/du$, pass through zero (twice), with $dg(u)/du>0$ between the minimum and maximum. The separation condition is discussed in Paper~2 with the background and the beam distributions considered as separated if for $ \varepsilon_n \ll 1 $ we have $ \gamma_{\rm b} \gtrsim \gamma_{\rm b, min} $ with
\begin{equation}\label{eq:gammabmin}
    \gamma_{\rm b, min} \approx
    7.8(1/\varepsilon_n\varepsilon_\rho\varepsilon_K)^{0.076}(1/\rho_0)^{1.07}.
\end{equation}
A rough estimate is $ \gamma_{\rm b, min} \sim 10\av{\gamma}_0 $ for the relevant parameter values. Over the region of interest, $ 1 > \zeta > \zeta_{\rm m0} $, where $ \zeta_{\rm m0} $ is the phase at which the RPDF $ \zeta^2 \Re W_0(\zeta) $ has a turning point, the fractional absorption $ \overline{\Gamma}_L(\zeta) $ is negative for (Paper~2)
\begin{equation}\label{eq:gamma12}
    \gamma_2 \lesssim \gamma_\phi \lesssim \gamma_1,
\end{equation}
where $ \gamma_1 \approx \gamma_{\rm b} $ and $ \gamma_2 $ corresponds to the phase where $ dg(u)/d\beta\big|_{\beta = \zeta} = 0 $ with $ \left.g(u)\right|_{\beta = \zeta} $ a local minimum and $ \sgn{(\zeta)} \left.dg(u)/d\beta\right|_{\beta = \zeta} > 0 $ over $ \gamma_2 \lesssim \gamma_\phi \lesssim \gamma_1 $. Therefore, as expected, negative absorption is possible for well-separated distributions over the phase range where $ \sgn{(\zeta)} \left.dg(u)/d\beta\right|_{\beta = \zeta} > 0 $.

\begin{figure}
\centering
\psfragfig[width=1.0\columnwidth]{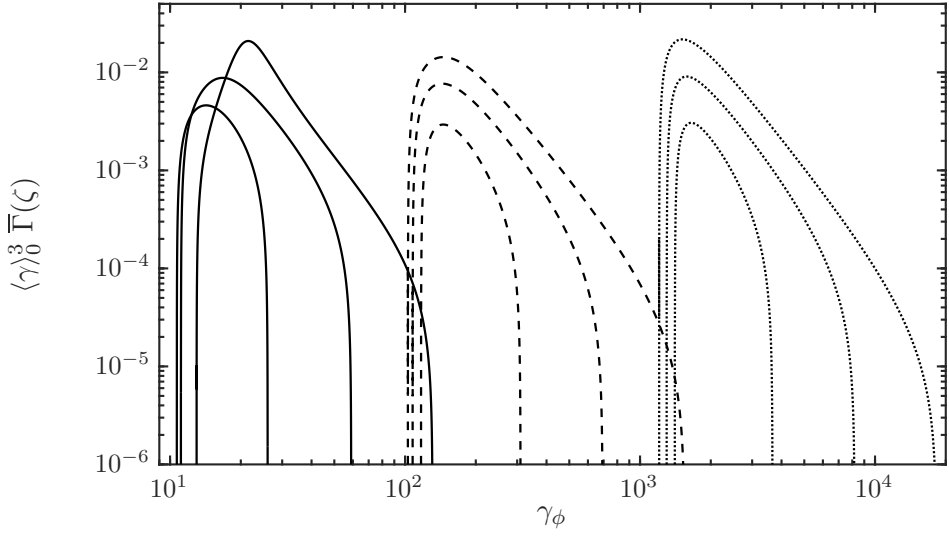}
 \caption{Plots of the absolute value of the negative region of the fractional absorption coefficient $ \overline{\Gamma}_L(\zeta) $, scaled by $ \av{\gamma}_0^3 $, as a function of $ \gamma_\phi $ for $ \rho_0 = 1 $ (solid), $ \rho_0 = 0.1 $ (dashed) and $ \rho_0 = 0.01 $ (dotted). For each $ \rho_0 $ we use $\varepsilon_\rho = 1 $, $ \varepsilon_n = 10^{-3} $ and $ \gamma_{\rm b}/\gamma_{\rm b, min}  = 2 $ (bottom), 3.2 (middle) and 5 (top). The right footpoint of each curve corresponds to $ \gamma_\phi \approx \gamma_{\rm b} $.}
 \label{fig:fabs_n}  
\end{figure}

In Figure~\ref{fig:fabs_n} we show plots of the absolute value of the negative region of the fractional absorption coefficient $ \overline{\Gamma}_L(\zeta) $, scaled by $ \av{\gamma}_0^3 $, as a function of $ \gamma_\phi $ for $ \rho_0 = 1 $ (solid), $ \rho_0 = 0.1 $ (dashed) and $ \rho_0 = 0.01 $ (dotted). For all values of $ \rho_0 $ we use $\varepsilon_\rho = 1 $, $ \varepsilon_n = 10^{-3} $ and $ \gamma_{\rm b}/\gamma_{\rm b, min}  = 2 $ (bottom), 3.2 (middle) and 5 (top). The right footpoint of each curve corresponds to $ \gamma_\phi \approx \gamma_1 \approx \gamma_{\rm b} $ and the left hand corresponds to $ \gamma_\phi \approx \gamma_2 $ as defined in~\eqref{eq:gamma12}. It is evident that the region over which $ \overline{\Gamma}_L(\zeta) $ is negative widens as $ \gamma_{\rm b} $ increases, however, the full width at half maximum (FWHM) remains relatively unaffected. The maximum negative value of $ \overline{\Gamma}_L(\zeta) $ occurs at $ \gamma_\phi \approx (12\pm4)\av{\gamma}_0 \sim \gamma_{\rm b, min} $ which is at higher phase velocity than the peak of $ \zeta^2\Re W_0(\zeta) $ at $ \zeta = \zeta_{\rm m0} $ corresponding to $ \gamma_\phi \approx \gamma_{\rm m0} \approx 6.2\av{\gamma}_0 $, where $ \gamma_{\rm m0} = \left.\gamma\right|_{\beta = \zeta_{\rm m0}}$.

\begin{figure}
\centering
\psfragfig[width=1.0\columnwidth]{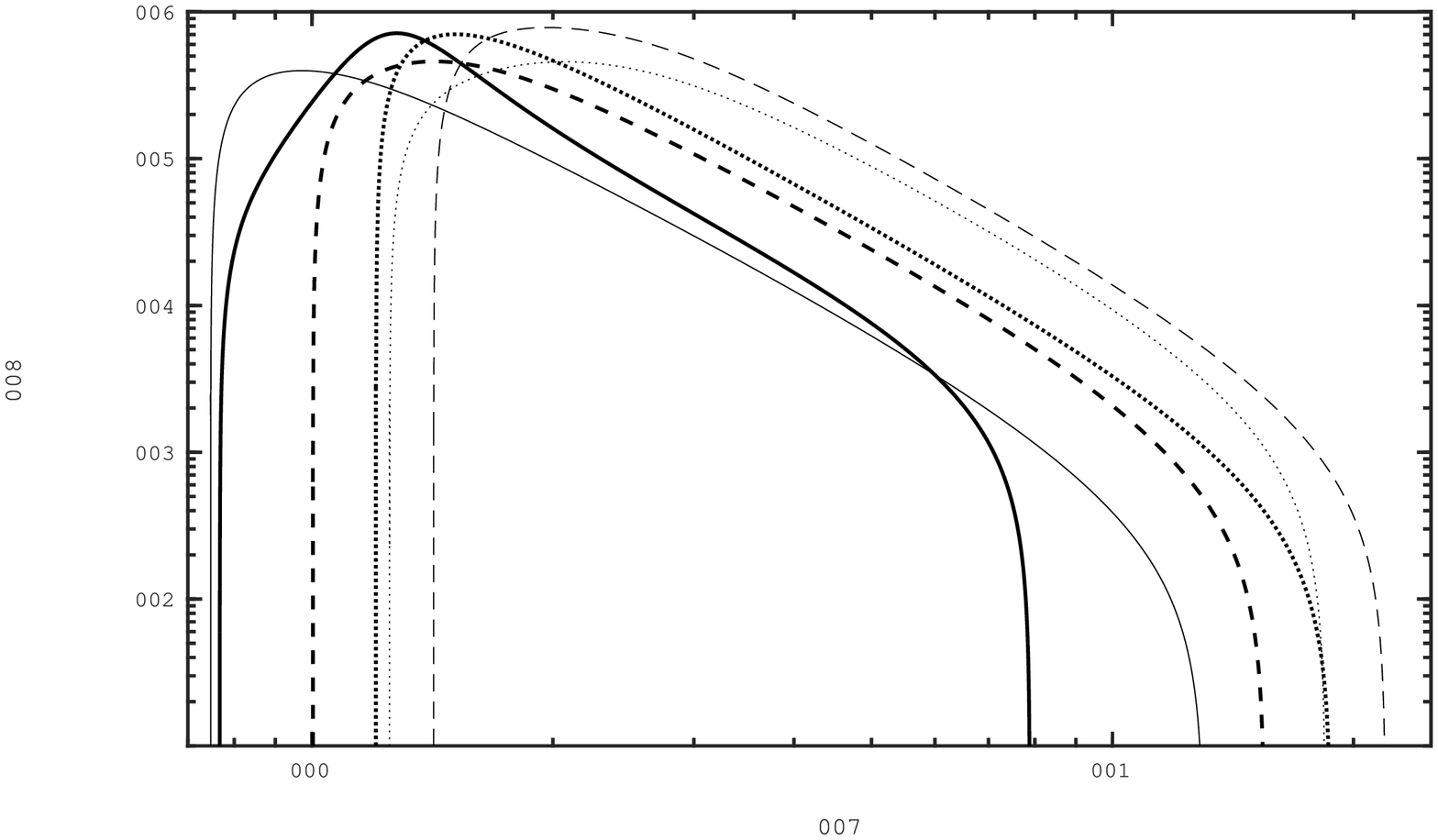}
 \caption{Plots of the absolute value of the negative region of $ \overline{\Gamma}_L(\zeta) $, scaled by $ \av{\gamma}_0^3/\varepsilon_\rho^{3/2}\varepsilon_n^{1/2} $, as a function of $ \gamma_\phi $, scaled by $ 1/\av{\gamma}_0 $, for $ \gamma_{\rm b}/\gamma_{\rm b, min}  = 3.2 $ and default values $ (\rho_0, \varepsilon_\rho, \varepsilon_n) = (0.1, 1, 10^{-3}) $. We vary $ \rho_0 \to 1 $ (solid), $ \rho_0 \to 0.01 $ (dotted), $ \varepsilon_n \to 10^{-4} $ (thin-dotted), $ \varepsilon_n \to 10^{-2} $ (thin-solid) and $ \varepsilon_n \to 0.1 $ (thin-dashed).}
 \label{fig:fabs_scaling}  
\end{figure}

In Figure~\ref{fig:fabs_scaling} we show plots of the absolute value of the negative region of $ \overline{\Gamma}_L(\zeta) $, scaled by $ \av{\gamma}_0^3/\varepsilon_\rho^{3/2}\varepsilon_n^{1/2} $, as a function of $ \gamma_\phi $, scaled by $ 1/\av{\gamma}_0 $, for $ \gamma_{\rm b}/\gamma_{\rm b, min}  = 3.2 $ and default values $ (\rho_0, \varepsilon_\rho, \varepsilon_n) = (0.1, 1, 10^{-3}) $. We vary $ \rho_0 \to 1 $ (solid), $ \rho_0 \to 0.01 $ (dotted), $ \varepsilon_n \to 10^{-4} $ (thin-dotted), $ \varepsilon_n \to 10^{-2} $ (thin-solid) and $ \varepsilon_n \to 0.1 $ (thin-dashed). This shows that the location of the peak roughly scales with $ 1/\av{\gamma}_0 $ and its height with $ \av{\gamma}_0^3/\varepsilon_\rho^{3/2}\varepsilon_n^{1/2} $.

\subsection{Penrose criterion}
\label{sec:Penrose}

In Appendix~\ref{app:Penrose} we derive a necessary and sufficient condition for temporal kinetic plasma instability ($\Im \omega > 0 $): the solution $ \omega^2 = \omega_{\rm p0}^2 z^2 W(z) $ to the dispersion equation $ K_{33}(\omega, k_\parallel) = 0 $ corresponds to a growing mode provided that $ \Re z^2 W(z) $ is positive when $ \Im z^2 W(z) $ switches sign from negative to positive at $ z = \zeta_2 $, where $ \zeta_2 $ corresponds to the phase at which $ \left.g(u)\right|_{\beta = \zeta_2} $ is a minimum with $ \left.dg(u)/d\beta\right|_{\beta = \zeta_2} = 0 $. When the separation condition is satisfied, Penrose criterion guarantees the existence of kinetic instability.

\subsection{Bandwidth of growing waves}

A restriction on the maser version of any instability is that the growth rate must be smaller than the bandwidth of the growing waves. This condition is required for the random phase approximation to be valid, with the bandwidth determining the rate of phase mixing of the growing waves. If the growth rate were to exceed the bandwidth of the growing waves, the initial phase would be remembered, and a phase-coherent wave would grow. The bandwidth of the growing waves provides a maximum value for the growth rate of the maser instability.  

In estimating  the bandwidth of the growing L~mode waves in a pulsar plasma, it is not obvious how one can appeal to the analogy with the non-relativistic thermal case, or its relativistic generalization to a Gaussian distribution. This is because the dispersion relation $\omega=\omega_{L0}(\zeta)$ is a rapidly varying function of $\zeta$ for the values of interest between the maximum of the RPDF $ \zeta^2\Re W_0(\zeta) $ at $\zeta=\zeta_{\rm m0} $ and the light line $\zeta=1$. This large rate of change of frequency with phase speed needs to be taken into account in estimating the bandwidth. A small range $\Delta \zeta$ corresponds to a very small fractional bandwidth. The range $ \Delta \zeta $ over which the fractional absorption $ \overline{\Gamma}_L(\zeta) $ is negative determines the bandwidth of the growing wave $ \Delta\omega $ through $ \Delta\omega = \Delta \zeta\, \left. d\omega_L(\zeta)/d\zeta\right|_{\zeta = \zeta_0} + \mathcal{O}\left((\Delta \zeta)^2\right) $ for some central $ \zeta_0 $. We estimate $ \Delta \zeta $ by the full width at half maximum (FWHM) of the negative region of the fractional absorption coefficient $ \overline{\Gamma}_L(\zeta) $ with $ \zeta_0 = \zeta_{\rm ma} $ corresponding to the phase at which the magnitude of the negative absorption coefficient is a maximum. The fractional bandwidth is then given by
\begin{equation}\label{eq:bw}
    \overline{\Delta\omega}_{\rm bw}
    = \left(\frac{\Delta\omega}{\omega_L(\zeta_{\rm ma})}\right)_{\rm bw}
    \approx \left.\frac{d\left[\zeta^2\Re W(\zeta)\right]/d\zeta}{2\zeta^2\Re W(\zeta)}\right|_{\zeta = \zeta_{\rm ma}} \Delta \zeta
    = \left[1 - \frac{1}{2R_L(\zeta_{\rm ma})}\right]\Delta \zeta.
\end{equation}

\begin{figure}
\centering
\psfragfig[width=1.0\columnwidth]{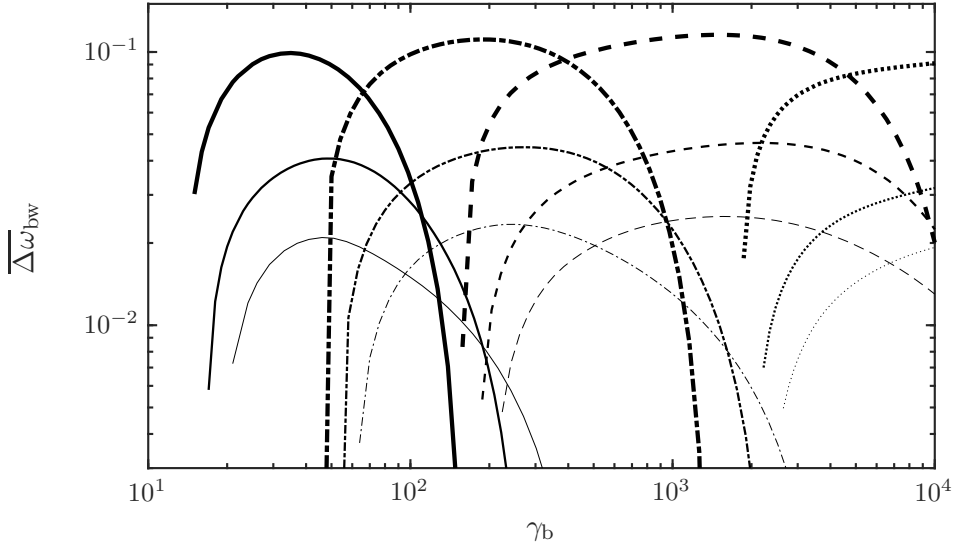}
 \caption{Plots of the fractional bandwidth $ \overline{\Delta\omega}_{\rm bw} $ over $ \gamma_{\rm b} $ for $ \rho_0 = 1 $ (solid), 0.32 (dash-dotted), 0.1 (dashed) and 0.01 (dotted). For each value of $ \rho_0 $ we use $ \varepsilon_\rho = 1 $ and $ \varepsilon_n = 10^{-3} $ (thick), $ 10^{-4} $ (medium) and $ 10^{-5} $ (thin).}
 \label{fig:fracBandwidth} 
\end{figure}

In Figure~\ref{fig:fracBandwidth} we show plots of the fractional bandwidth $ \overline{\Delta\omega}_{\rm bw} $ over $ \gamma_{\rm b} $ for $ \rho_0 = 1 $ (solid), 0.32 (dash-dotted), 0.1 (dashed) and 0.01 (dotted). For each value of $ \rho_0 $ we use $ \varepsilon_\rho = 1 $ and $ \varepsilon_n = 10^{-3} $ (thick), $ 10^{-4} $ (medium) and $ 10^{-5} $ (thin). Using $ \varepsilon_n = 10^{-3}, 10^{-4}, 10^{-5} $, for $ \rho_0 = 1 $ the fractional bandwidth is a maximum at $ \gamma_{\rm b} \approx 35, 49, 46 $, for $ \rho = 0.32 $ the maxima are at $ \gamma_{\rm b} \approx (1.9, 2.8, 2.4)\times10^2$, and for $ \rho_0 = 0.1 $ the maxima are at $ \gamma_{\rm b} \approx (1.6, 2.1, 1.6)\times10^3$, respectively. The maximum value of the fractional bandwidth is given by
\begin{equation}
    \overline{\Delta\omega}_{\rm bw,~max} \approx 2\times10^{-3}\varepsilon_n^{-1/3}.
\end{equation}

For the purpose of discussing the kinetic instability we are interested in the ratio of the growth rate to the bandwidth of the growing wave
\begin{equation}\label{eq:ratios}
    \varepsilon_L = \frac{\overline{\Gamma}_L(\zeta_{\rm ma})}{\overline{\Delta\omega}_{\rm bw}}.
\end{equation}
For maser growth through kinetic instability to be possible we require that the growth rate is smaller than the bandwidth of the growing wave so that phase mixing can occur. This is equivalent to requiring that $ \varepsilon_L < 1 $.

\begin{figure}
\centering
\psfragfig[width=1.0\columnwidth]{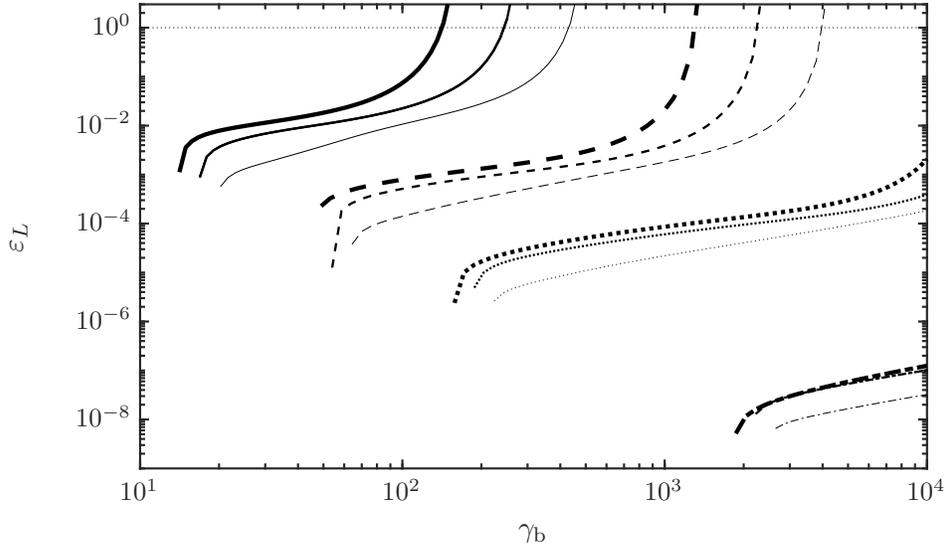}
 \caption{Plot $ \varepsilon_L $ over $ \gamma_{\rm b} $ for $ \rho_0 = 1 $ (solid), 0.32 (dashed), 0.1 (dotted) and 0.01 (dash-dotted). For each value of $ \rho_0 $ we use $ \varepsilon_\rho = 1 $ and $ \varepsilon_n = 10^{-3} $ (thick), $ 10^{-4} $ (medium) and $ 10^{-5} $ (thin). The thin dotted horizontal line indicates the threshold $ \varepsilon_L = 1 $.}
 \label{fig:epsL}  
\end{figure}

In Figure~\ref{fig:epsL} we plot $ \varepsilon_L $ over $ \gamma_{\rm b} $ for $ \rho_0 = 1 $ (solid), 0.32 (dashed), 0.1 (dotted) and 0.01 (dash-dotted). For each value of $ \rho_0 $ we use $ \varepsilon_\rho = 1 $ and $ \varepsilon_n = 10^{-3} $ (thick), $ 10^{-4} $ (medium) and $ 10^{-5} $ (thin). The thin dotted horizontal line indicates the threshold $ \varepsilon_L = 1 $. For all values of $ \rho_0 $, each curve starts at $ \gamma_{\rm b} \approx \gamma_{\rm b, min} $. For $ \rho_0 = 1, 0.32 $ we have $ \varepsilon_L > 1 $ at $ \gamma_{\rm b} \gtrsim (1.4, 2.5, 4.2) \times 10^2 $ and $ \gamma_{\rm b} \gtrsim (1.3, 2.2, 4) \times 10^3 $, respectively, for $ \varepsilon_n = (10^{-3}, 10^{-4}, 10^{-5}) $. The streaming Lorentz factor $ \gamma_{\rm b} $ for which $ \varepsilon_L > 1 $ roughly scales as $ \sim (25 {\rm -} 23)/\varepsilon_n^{1/4}\rho_0^2 $ for $ \rho_0 = 1 {\rm -} 0.32 $. Furthermore, the curves corresponding to $ \rho_0 = 1,0.32 $ do not extend to $ \gamma_{\rm b} = 10^4 $ because growth is suppressed for $ \gamma_{\rm b} \gtrsim (1.7, 2.9, 5) \times10^2 $ and $ \gamma_{\rm b} \gtrsim (1.4, 2.4, 4.3) \times10^3 $, respectively, for $ \varepsilon_n = (10^{-3}, 10^{-4}, 10^{-5}) $. 

\subsection{Comparison with the results of ELM}

ELM considered wave growth in a pulsar plasma assuming a relativistically streaming Gaussian beam. These authors estimated both the growth rate and the bandwidth of the growing waves, and argued that the requirement that the bandwidth exceed the growth rate is so severe that it cannot be satisfied for plausible parameters in a pulsar plasma. Here we apply the same argument to a Lorentz-transformed J\"uttner distribution. 

For the relativistically streaming Gaussian distribution (\ref{Gaussian}) the bandwidth of the growing waves was estimated by ELM to be $\Delta\omega_{\rm bw}\approx|k_\parallel|c\Delta\beta$, where $\Delta\beta \approx \beta\Delta\gamma/\gamma^3$ is the range of speeds over which the resonance applies. An analogy between  a relativistically streaming Gaussian and a streaming Maxwellian allows this width to be identified as $\Delta\gamma\to\gamma_{\rm Tb}$ around $\gamma=\gamma_{\rm b}$, giving $\Delta\omega_{\rm bw}\approx|k_\parallel|c\gamma_{\rm Tb}/\gamma_{\rm b}^3$. As in the nonrelativistic case, the spread in phase speeds is identified as $\Delta \zeta=\Delta\beta$ around the resonant value $\zeta=\beta_{\rm b}$, corresponding to a spread $\Delta\gamma_\phi$ about $\gamma_\phi=\gamma_{\rm b}$, with $\Delta \zeta=\Delta\gamma_\phi/\gamma_\phi^3$. ELM compared this bandwidth with the maximum growth rate for the kinetic instability due to the relativistically streaming Gaussian distribution, and found that the growth rate greatly exceeds the bandwidth for plausible parameters. We note that, although ELM took the dispersion relation for waves in pulsar plasma into account elsewhere, they did not take it into account explicitly in estimating this bandwidth.

ELM suggested that plausible parameters are $ n_1/n_0 \approx 10^{-3} $ which corresponds to $ \varepsilon_n = n_1/\gamma_{\rm b}n_0 \approx 10^{-3}/\gamma_{\rm b} $. The minimum separation streaming Lorentz factors required for $ \rho_0 = 1, 0.1, 0.01 $ are $ \gamma_{\rm b, min} \sim 10, 10^2, 10^3 $, respectively, which implies corresponding $ \varepsilon_n $ of $ \sim 10^{-4}, 10^{-5}, 10^{-6} $. The results of \cite{AE02} suggest that particle distributions (background or beam) stream with Lorentz factors $ \sim 10^2{\rm -}10^{3} $ in the pulsar frame. If one assumes that, in the pulsar frame, the rest frame of the background distribution streams with $ \gamma_{\rm s} \sim 10^2 {\rm -} 10^3 $ then the rest frame of the beam must stream with Lorentz factor $ \gamma_{\rm r} \sim 10^{3} {\rm -} 10^4 $ for $ \rho_0 = 1 $; $ \gamma_{\rm r} \sim 10^{4} {\rm -} 10^5 $ for $ \rho_0 = 0.1 $; and $ \gamma_{\rm r} \sim 10^{5} {\rm -} 10^6 $ for $ \rho_0 = 0.01 $. The peak of negative absorption is at $ \gamma_\phi \sim 10\av{\gamma}_0 $ which implies that for maximum wave growth we want $ \gamma_{\rm b} \sim 10\av{\gamma}_0 $; this roughly corresponds to $ \gamma_{\rm b, min} $. We may then write, for maximum fractional growth rate, $ \gamma_{\rm r} \sim (10^{3}{\rm -}10^4) \av{\gamma}_0 $ using $ \gamma_{\rm r} = \gamma_{\rm s}\gamma_{\rm b} $.

The maximum fractional growth rate may be approximated from Figure~\ref{fig:fabs_scaling} as
\begin{equation}\label{eq:maxneg2}
    \left|\overline{\Gamma}_L(\zeta)\right|_{\rm max} \sim \frac{\varepsilon_n^{1/2}\varepsilon_\rho^{3/2}}{2 \av{\gamma}_0^3}.
\end{equation}
For $ \varepsilon_\rho = 1 $, $ \rho_0 = 1, 0.1, 0.01 $ and corresponding $ \varepsilon_n = 10^{-4}, 10^{-5}, 10^{-6} $ we have, respectively,
\begin{equation}\label{eq:maxneg3}
    \left|\overline{\Gamma}_L(\zeta)\right|_{\rm max} \sim 10^{-3}, 10^{-6}, 10^{-11}.
\end{equation}

We conclude that wave growth through kinetic (or maser) instability is possible (in principle). This is different from the conclusion of ELM; this difference reflects the difference between the relativistically streaming Gaussian distribution that they assumed and the (much broader) Lorentz-transformed distribution that we assume, and for which the bandwidth is larger and the growth rate is smaller.

\section{Reactive version of weak-beam instability}
\label{sect:reactive}

In this section we consider reactive versions of the weak-beam instability in a pulsar plasma. A reactive instability is identified as one of a pair of complex conjugate solutions of a real dispersion equation. Various approximations need to be made to reduce the dispersion equation to a real equation, similar to the cold-plasma form of the dispersion equation, in order to derive such analytic solutions. In the case of a nonrelativistic thermal distribution, or a relativistically streaming Gaussian distribution, for the beam, a cold-plasma-like form is obtained by assuming the bandwidth is much less than the growth rate, effectively ignoring the thermal spread. However, more questionable assumptions are needed to reduce the dispersion equations for J\"uttner distributions to a cold-plasma-like form. Numerical calculations are needed to determine the validity of analytic solutions derived in this way. We find that our numerical calculations do not justify the analytic results for parameters thought plausible for a relativistically streaming J\"uttner distribution in a pulsar magnetosphere. This leads us to suggest that there may be no reactive instability for a model based on J\"uttner distributions.

\subsection{Reduction of the dispersion equation}

In treating a reactive instability, one neglects the dissipative part of the response function, which plays a central role in the treatment of a kinetic instability. This corresponds to neglecting the dissipative part of the RPDF and the imaginary parts of $\omega$ and $k_\parallel$, so that the response function \eqref{eq:K33r0} reduces to
\begin{equation}\label{eq:K33r1}
    K_{33}(\omega, k_\parallel)
        = 1 - \omega_{\rm p0}^2\frac{\zeta^2 \Re W_0(\zeta)}{\omega^2} - \varepsilon_n \omega_{\rm p0}^2\frac{\zeta''^2 \Re W''_1(\zeta'')}{\gamma_{\rm b}^2(\omega - \beta_{\rm b}ck_\parallel)^2}.
\end{equation}
In treating reactive instability we assume $ k_\parallel = \Re k_\parallel $ and comment on the spatial growth in the next section. To identify a reactive instability one needs to approximate the dispersion equation $ K_{33}(\omega, k_\parallel) = 0 $, implied by setting \eqref{eq:K33r1} to zero, by a polynomial in $ \omega $ with real coefficients. The complex solutions then must appear in complex conjugate pairs, one of which describes an intrinsically growing wave. 

The simplest cases of reactive instabilities are derived assuming a cold-plasma model, such that the spread in particle energies is neglected. For a cold background, $ \rho_0 = \infty $, and cold beam, $ \rho_1 = \infty $, one has $ \zeta^2 \Re W_0(\zeta) \to 1 $ and $ \zeta''^2 \Re W''_1(\zeta'') \to 1 $ so that~\eqref{eq:K33r1} reproduces the well-known cold plasma form for the dispersion equation for a relativistic cold weak-beam system:
\begin{equation}
\label{eq:cold}
    1 - \frac{\omega_{\rm p0}^2}{\omega^2} - \frac{\omega_{\rm p1}^2}{\gamma_{\rm b}^3(\omega - \beta_{\rm b} ck_\parallel)^2} = 0,
\end{equation}
which is a quartic equation in $\omega$. A complication in reducing~\eqref{eq:K33r1} to this form is  that $ \zeta $ and $ \zeta'' $ are implicit functions of $ \omega $. To proceed analytically, we ignore this complication, assuming that $ \zeta $ and $ \zeta'' $ are constants. In our numerical results we solve~\eqref{eq:K33r1} directly without additional assumptions.
 
A justification for assuming that $ \zeta $ and $ \zeta'' $ are constants is that they are both close to unity, suggesting that one may approximate them by unity, giving
\begin{equation}\label{eq:omeL01_approx}
    \zeta^2 \Re W_0(\zeta) \to \Re W_0(1) = 2\av{\gamma}_0 - \av{1/\gamma}_0,\quad
    \zeta''^2 \Re W''_1(\zeta'') \to \Re W''_1(1) = 2\av{\gamma''}''_1 - \av{1/\gamma''}''_1,
\end{equation}
where the right hand expressions were derived in Paper~1. For $ \{\rho_0, \rho_1\} \ll 1 $ we may write $\Re W_0(1) \approx 2\av{\gamma}_0 $ and $ \Re W''_1(1) = 2\av{\gamma''}''_1 $. With the implicit dependence on $\omega$ through $\zeta$ and $\zeta''$ neglected in this way, the dispersion equation~\eqref{eq:K33r1} becomes cold-plasma-like and has the form
\begin{equation}\label{eq:K33r1_approx}
    K_{33}(\omega, k_\parallel)
        = 1 - \frac{\omega_{L0}^2(1)}{\omega^2} - \frac{\omega''^2_{L1}(1)}{\gamma_{\rm b}^2(\omega - \beta_{\rm b}ck_\parallel)^2},
\end{equation}
where $ \omega^2_{L0}(\zeta) = \omega_{\rm p0}^2 \zeta^2 \Re W_0(\zeta) $ and $ \omega''^2_{L1}(\zeta'') = \varepsilon_n \omega_{\rm p0}^2 \zeta''^2 \Re W''_1(\zeta'') $. With these assumptions, dispersion equation~\eqref{eq:K33r1_approx} becomes a polynomial of degree four in $ \omega $ with real coefficients. 

\begin{figure}
\centering
\psfragfig[width=1.0\columnwidth]{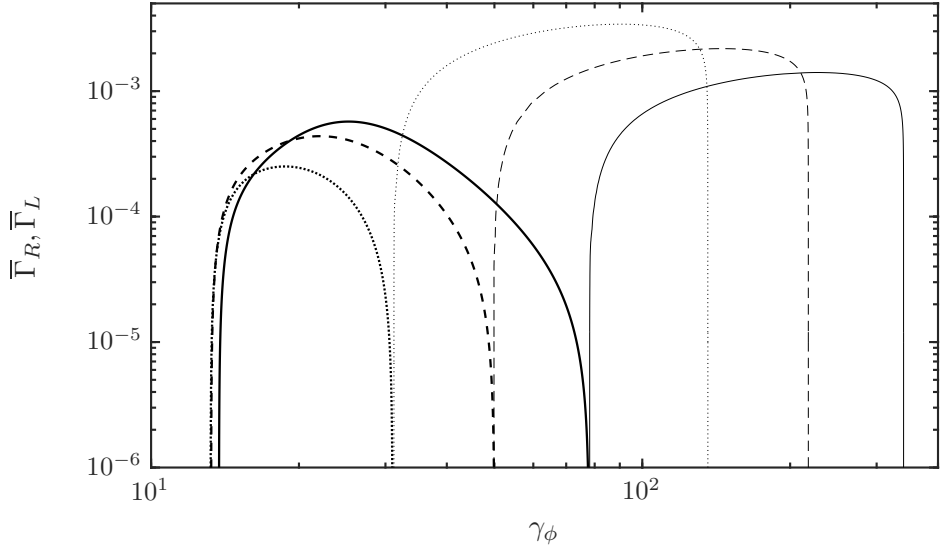}
 \caption{Plots of the fractional growth rate for reactive instability $ \overline{\Gamma}_R $ (thin curves) and for kinetic instability $ \overline{\Gamma}_L $ (thick curves) for $ \rho_0 = 1 $, $ \varepsilon_\rho = 1 $, $ \varepsilon_n = 10^{-4} $, and $ \gamma_{\rm b}/\gamma_{\rm b, min} = 2 $ (dotted), 3.2 (dashed) and 5 (solid).}
 \label{fig:Reactive}  
\end{figure}

\subsubsection{Numerical solutions of the dispersion equation}

We numerically solve~\eqref{eq:K33r1}, treating $ \zeta $ as a function of $ \omega_r $, for $ \omega = \omega_r + i\omega_i $ and define the fractional growth rate of (temporal) reactive instability as
\begin{equation}
    \overline{\Gamma}_R = 2\frac{\omega_i}{\omega_r}.
\end{equation}
Figure~\ref{fig:Reactive} shows plots of the fractional growth rate for reactive instability $ \overline{\Gamma}_R $ (thin curves) and for kinetic instability $ \overline{\Gamma}_L $ (thick curves) for $ \rho_0 = 1 $, $ \varepsilon_\rho = 1 $, $ \varepsilon_n = 10^{-4} $, and $ \gamma_{\rm b}/\gamma_{\rm b, min} = 2 $ (dotted), 3.2 (dashed) and 5 (solid). It is clear that reactive instability and kinetic instability apply over different ranges of phase velocity $ \zeta $; however, they overlap when plotted over wavenumber $ k_\parallel $ which will become apparent below. Reactive instability grows faster, and applies over a wider phase velocity range and closer to the light line $ \zeta = 1 $ than kinetic instability. 

Reactive instability applies over $ \gamma_{\rm b} \lesssim \gamma_\phi \lesssim \left.\gamma\right|_{\beta = \zeta_{01}} $ where $ \zeta_{01} $ is where $ \zeta^2 \Re W_1(\zeta) $ is zero. In the rest frame of the beam $ \zeta''_{01} \approx 1 - \alpha_1 \rho_1 $ with $ \alpha_1 \approx 0.1 $ for $ \rho_1 = 1 $ and $ \alpha_1 \approx 0.144 $ for $ \rho_1 \ll 1 $ (Paper~1). This corresponds to $ \left.\gamma\right|_{\beta = \zeta_{01}} \approx 4.4 \gamma_{\rm b} $ for $ \rho_1 = 1 $ and $ \left.\gamma\right|_{\beta = \zeta_{01}} \approx 3.5 \gamma_{\rm b} \av{\gamma}_0/\varepsilon_\rho $ for $ \rho_1 \ll 1 $.

\subsection{Resonant reactive instability}

The most familiar form of a reactive weak-beam instability is the solution of a cubic equation. To derive this case, we treat the contribution of the beam as a perturbation to that of the background so that to zeroth order the frequency is given by $ \omega_{L0}(1) $. It is convenient to assume $ k_\parallel = \Re k_\parallel $ and introduce
\begin{equation}
    \delta\omega = \omega - \omega_{L0}(1),\quad
    \delta\omega_0 = \omega_{L0}(1) - k_\parallel c\beta_{\rm b},
    \label{reac1}
\end{equation}
with $ |\delta\omega/\omega_{L0}(1)| \ll 1 $ so that $ K_{33}(\omega, k_\parallel) = 0 $ in~\eqref{eq:K33r1_approx} may be approximated as
\begin{equation}\label{eq:K33r2}
    \overline{\delta\omega} (\overline{\delta\omega} + \overline{\delta\omega}_0)^2 
        \approx 1,
\end{equation}
with
\begin{equation}
    \frac{\delta\omega}{\overline{\delta\omega}}
        = \frac{\delta\omega_0}{\overline{\delta\omega}_0}
        = [\omega_{L0}(1)\omega''^2_{L1}(1)/2\gamma_{\rm b}^2]^{1/3}.
\end{equation}

The maximum growth rate is obtained when $ |\delta\omega_0| \ll |\delta\omega| $ so that~\eqref{eq:K33r2} reduces to $ (\overline{\delta\omega})^3 \approx 1 $ with solutions
\begin{equation}
    \overline{\delta\omega} = 1, \quad
    \overline{\delta\omega} = -(1 \pm i\sqrt{3})/2.
\end{equation}
The growing solution, referred to as the resonant reactive instability, is
\begin{equation}
    \omega = \omega_{L0}(\zeta)\left[1 - (1 - i\sqrt{3})\left(\frac{\omega''^2_{L1}(1)}{16\gamma_{\rm b}^2\omega_{L0}^2(1)}\right)^{1/3}\right].
\end{equation}
The fractional growth rate of the resonant reactive instability is then given by
\begin{equation}\label{eq:ReactiveGrowthRate}
    \overline{\Gamma}_{\rm rr}
        = 2\frac{\omega_i}{\omega_r}
        \approx \sqrt{3}\left(\frac{\varepsilon_n\av{\gamma''}''_1}{2\gamma_{\rm b}^2\av{\gamma}_0}\right)^{1/3}
        = \frac{\sqrt{3}}{\gamma_{\rm b}}\left(\frac{n_1\av{\gamma''}''_1}{2n_0\av{\gamma}_0}\right)^{1/3},
\end{equation}
where we use $ \omega^2_{L0}(1) \approx 2\omega_{\rm p0}^2 \av{\gamma}_0 $, $ \omega''^2_{L1}(1) \approx 2\varepsilon_n\omega_{\rm p0}^2\av{\gamma''}''_1 $ and assume $ \varepsilon_n\av{\gamma''}''_1/2\gamma_{\rm b}^2\av{\gamma}_0 \ll 1 $. The fractional growth rate of the resonant reactive instability decreases with increasing streaming Lorentz factor $ \gamma_{\rm b} $ and increases with $ \varepsilon_n $. The fractional growth rate does not depend independently on the temperatures of the two distributions, described by the mean Lorentz factors, $ \av{\gamma}_0 $ and $ \av{\gamma''}''_1 $, but only on the ratio of these quantities. This lack of dependence on temperature is not surprising as we effectively approximate the background and the beam as cold distributions through ignoring the implicit dependence of $ \zeta $ and $ \zeta'' $ on $ \omega $.

The analytical approximation~\eqref{eq:ReactiveGrowthRate} is supported by numerical solutions of~\eqref{eq:K33r1} where $ \zeta $ and $ \zeta'' $ are treated as functions of $ \omega $. The thin-dotted curve in Figure~\ref{fig:Reactive} corresponds to $ \rho_0 = 1 $, $ \varepsilon_\rho = 1 $, $ \varepsilon_n = 10^{-4} $ and $ \gamma_{\rm b} \approx 32 $ with peak value $ \overline{\Gamma}_R \approx 3.4\times10^{-3} $. For these parameter values we obtain $ \overline{\Gamma}_{\rm rr} \approx 6.3\times10^{-3} $ using~\eqref{eq:ReactiveGrowthRate}.

\subsection{Non-resonant reactive instability}

In~\eqref{eq:K33r1_approx}, contribution of the third term is significant when $ \omega - \beta_{\rm b} ck_\parallel $ is small since $ \varepsilon_n \ll 1 $. Introducing
\begin{equation}
    \delta\omega_1 = \omega - \beta_{\rm b}ck_\parallel,
\end{equation}
with $ |\delta\omega_1/\beta_{\rm b} ck_\parallel| \ll 1 $ allows us to write~\eqref{eq:K33r1_approx} as
\begin{equation}
    K_{33}(\omega, k_\parallel)
        \approx 1 - \frac{\omega_{L0}^2(1)}{(\beta_{\rm b}ck_\parallel)^2} - \frac{\omega''^2_{L1}(1)}{\gamma_{\rm b}^2(\delta\omega_1)^2},
\end{equation}
with $ K_{33}(\omega, k_\parallel) = 0 $ having solutions
\begin{equation}\label{eq:omegann}
    \omega = \beta_{\rm b}ck_\parallel \pm \frac{\beta_{\rm b}ck_\parallel\omega''_{L1}(1)/\gamma_{\rm b}}{[(\beta_{\rm b}ck_\parallel)^2 - \omega_{L0}^2(1)]^{1/2}}.
\end{equation}
A growing mode ($ \omega_i > 0 $) exists only if $ \beta_{\rm b} c k_\parallel < \omega_{L0}(1) $. This condition agrees with our discussion that reactive instability applies over $ \gamma_{\rm b} \lesssim \gamma_\phi \lesssim \left.\gamma\right|_{\beta = \zeta_{01}} $ (see also Figure~\ref{fig:W0_W1_W} below). This is the non-resonant unstable beam mode~\citep{GGM02} with real frequency $ \omega_r \approx \beta_{\rm b} c k_\parallel $. The non-resonant fractional absorption coefficient is given by
\begin{equation}\label{eq:RGnr}
    \overline{\Gamma}_{\rm nr} 
        = 2\frac{\omega_i}{\omega_r}
        = \frac{2\omega''_{L1}(1)/\gamma_{\rm b}}{[\omega_{L0}^2(1) - (\beta_{\rm b}ck_\parallel)^2]^{1/2}}
        \approx \left(\frac{2}{\gamma_{\rm b}}\right)^{3/2}\left(\frac{n_1\av{\gamma''}''_1}{2n_0\av{\gamma}_0}\right)^{1/2}\left[1 - \frac{(\beta_{\rm b}ck_\parallel)^2}{2\av{\gamma}_0\omega_{\rm p0}^2}\right]^{-1/2}.
\end{equation}
Clearly~\eqref{eq:omegann} and~\eqref{eq:RGnr} break down as $ \beta_{\rm b} ck_\parallel \to \omega_{L0}(1) $. This limit corresponds to the resonant case as treated above where $ \omega_r \approx \omega_{L0}(1) $ instead of $ \omega_r \approx \beta_{\rm b}ck_\parallel $ in the non-resonant case. We may approximate the ratio of the resonant to non-resonant fractional growth rate as
\begin{equation}
    \frac{\overline{\Gamma}_{\rm rr}}{\overline{\Gamma}_{\rm nr}}
        \approx \sqrt{3}\left(\frac{\gamma_{\rm b}}{8}\right)^{1/2}\left(\frac{n_1\av{\gamma''}''_1}{2n_0\av{\gamma}_0}\right)^{-1/6},
\end{equation}
where we use $ \beta_{\rm b}ck_\parallel/\omega_{L0}(1) \ll 1 $ for the non-resonant case. The non-resonant fractional growth rate is smaller than that for the resonant reactive instability, but it applies over a wider frequency range, which tends to offset the smaller growth rate \citep{GMG02}. The argument is that the gain factor depends on the number of e-folding growth lengths which is limited by the changing value of $\omega_{\rm p0}$, and hence $\omega_{L0}(1)$, along the ray path: the growth rate (\ref{eq:RGnr}) allows the waves to grow over a much longer ray path than for the resonant reactive version.

\subsubsection{Comparison with ELM}
 
For comparison, the estimate made by ELM for a streaming relativistic Gaussian distribution, $\propto\exp[-(u-u_{\rm b})^2/\langle\gamma\rangle^2]$, may be written as
\begin{equation}
    2\left(\frac{\omega_i}{\omega_r}\right)_{\rm rr}
        \approx \frac{\sqrt{3}}{\gamma_{\rm b}}\frac{(n_1/n_0)^{1/3}}{10\langle\gamma\rangle_0}
        \sim \frac{10^{-3}}{\av{\gamma}_0^2},
\label{ELM2}
\end{equation}
with $ n_1/n_0 \sim 10^{-3} $ and where we use $ \gamma_{\rm b} \sim 10\av{\gamma}_0 $ as in~\eqref{eq:maxneg2}. We may estimate~\eqref{eq:ReactiveGrowthRate} as $ \overline{\Gamma}_{\rm rr} \sim 10^{-2}/\av{\gamma}_0 $ for these parameter values. The resonant reactive growth rate for a streaming Gaussian distribution, (\ref{ELM2}), is smaller than than for a J\"uttner distribution, (\ref{eq:ReactiveGrowthRate}), by a factor $ \sim 1/10\av{\gamma}_0 $.

\subsection{Does reactive instability occur?}

Our results appear to be implying that reactive instability occurs for J\"uttner distributions in a pulsar plasma. We first describe our search for reactive instability, and then discuss the interpretation and implications of our results.

\subsubsection{Search for reactive instability}
The procedure we have employed in this section is consistent with existing literature. However, when one solves the exact dispersion equation~\eqref{wb1} numerically using~\eqref{eq:Wcases}, without expanding in $ \{|\omega_i/\omega_r|, |k_i/k_r|\} \ll 1 $, only the kinetic solution appears. This discrepancy is resolved by critically examining the approximate form of the dispersion equation~\eqref{eq:K33r1} used in treating the reactive instability.

In~\eqref{eq:K33r1}, for $ k_\parallel = \Re k_\parallel $, we write $ z^2W_0(z) \approx \zeta^2\Re W_0(\zeta) $ assuming $ |\omega_i/\omega_r| \ll 1 $ in the rest frame of the background. Similarly, assuming $ |\omega''_i/\omega''_r| \ll 1 $ in the rest frame of the beam allows us to write $ z''^2W''_1(z'') \approx \zeta''^2\Re W''_1(\zeta'') $. As discussed in Paper~1, the assumption $ |\omega_i/\omega_r| \ll 1 $ ($ |\omega''_i/\omega''_r| \ll 1 $) is valid only for $ |\zeta| \gtrsim \zeta_{\rm m0} $ ($ |\zeta''| \gtrsim \zeta''_{\rm m1} $) where $ \zeta_{\rm m0} $ ($ \zeta''_{\rm m1} $) is the phase at which the RPDF $ \zeta^2W_0(\zeta) $ ($ \zeta''^2W''_1(\zeta'') $) has a positive turning point. Landau damping dominates for $ |\zeta| \lesssim \zeta_{\rm m0} $ ($ |\zeta''| \lesssim \zeta''_{\rm m1} $).

\begin{figure}
\centering
\psfragfig[width=1.0\columnwidth]{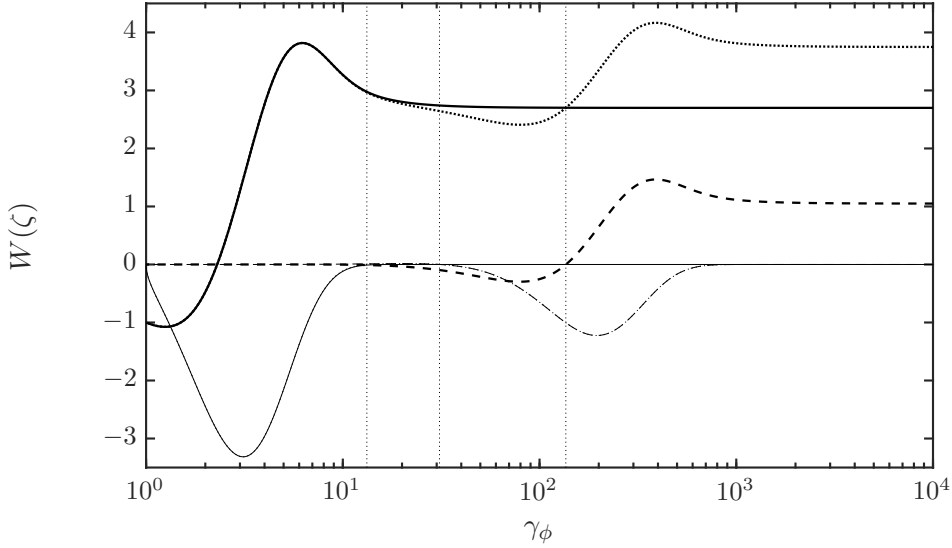}
 \caption{Plots of real (thick) and imaginary (thin) parts of $ W(\zeta) $ (dotted), $ W_0(\zeta) $ (solid) and $ \varepsilon_n\gamma_{\rm b} W_1(\zeta) $ (dashed) over $ \gamma_\phi $ for $ \rho_0 = 1 $, $\varepsilon_\rho = 1 $, $ \varepsilon_n = 10^{-4} $ and $ \gamma_{\rm b}/\gamma_{\rm b, min} = 2 $. The three vertical dotted lines indicate, from left to right, $ \gamma_\phi = \gamma_2, \gamma_1, \left.\gamma\right|_{\beta = \zeta_{01}} $, respectively, with $ \gamma_1 \approx \gamma_{\rm b} $.}
 \label{fig:W0_W1_W}  
\end{figure}

Figure~\ref{fig:W0_W1_W} shows plots of real (thick) and imaginary (thin) parts of $ W(\zeta) $ (dotted), $ W_0(\zeta) $ (solid) and $ \varepsilon_n\gamma_{\rm b} W_1(\zeta) $ (dashed) over $ \gamma_\phi $ for $ \rho_0 = 1 $, $\varepsilon_\rho = 1 $, $ \varepsilon_n = 10^{-4} $ and $ \gamma_{\rm b}/\gamma_{\rm b, min} = 2 $. The three vertical dotted lines indicate, from left to right, $ \gamma_\phi = \gamma_2, \gamma_1, \left.\gamma\right|_{\beta = \zeta_{01}} $, respectively, with $ \gamma_1 \approx \gamma_{\rm b} $, and $ \left.\gamma\right|_{\beta = \zeta_{01}} \approx 4.4 \gamma_{\rm b} $ for $ \rho_1 = 1 $ and $ \left.\gamma\right|_{\beta = \zeta_{01}} \approx 3.5 \gamma_{\rm b} \av{\gamma}_0/\varepsilon_\rho $ for $ \rho_1 \ll 1 $, as discussed above. The imaginary part of $ W(\zeta) $ follows $ \Im W_0(\zeta) $ for $ \gamma_\phi \lesssim \gamma_2 $ and $ \varepsilon_n\gamma_{\rm b} \Im W_1(\zeta) $ for $ \gamma_\phi \gtrsim \gamma_{\rm b} $. Kinetic instability applies over $ \gamma_1 \lesssim \gamma_\phi \lesssim \gamma_{\rm b} $ and reactive instability applies over $ \gamma_{\rm b} \lesssim \gamma_\phi \lesssim \left.\gamma\right|_{\beta = \zeta_{01}} $. Both kinetic and reactive instability apply over a region where $ \zeta > \zeta_{\rm m0} $, but for both instabilities one has $ \zeta'' < \zeta''_{\rm m1} $. This is not an issue when we are treating kinetic instability where the beam is a perturbation to the background. Furthermore, in treating kinetic instability we make the approximation $ z^2W(z) \approx \zeta^2 \Re W(\zeta) $, with $ W(z) $ defined in~\eqref{eq:omegawb}, for which indeed $ |\omega_i/\omega_r| \ll 1 $ over the region where kinetic instability applies, as seen in Figure~\ref{fig:W0_W1_W}. Over the region where reactive instability applies, the beam is not a perturbation to the background except near $ \gamma_\phi = \gamma_{\rm b} $ where non-resonant reactive instability applies. Also, $ |\omega_i/\omega_r| \ll 1 $ is not satisfied properly especially near $ \gamma_\phi = \left.\gamma\right|_{\beta = \zeta_{01}} $ where resonant reactive instability applies.

\subsubsection{Why is there no reactive instability?}

When damping/growth is weak we may write $ W(\zeta) $ using~\eqref{eq:Wdef} as
\begin{equation}\label{eq:Wweak}
    W(\zeta) = \frac{1}{n_0}\int_{-1}^1d\beta\, \frac{dg(u)/d\beta}{\beta - \zeta},
\end{equation}
for $ W(\zeta) = W_0(\zeta) + \varepsilon_n\gamma_{\rm b}W_1(\zeta) $ with $ g(u) $ given by~\eqref{eq:gdgdbeta}. Interaction of a wave with phase velocity $ \zeta $ and a particle with velocity $ \beta $ is resonant if in~\eqref{eq:Wweak} the Cenerkov resonance condition $ \beta - \zeta = 0 $ is satisfied; the interaction is non-resonant otherwise. Kinetic instability arises from resonant interaction over regions where $ \sgn{(\zeta)}\left.dg(u)/d\beta\right|_{\beta = \zeta} > 0 $ and Landau damping occurs when $ \sgn{(\zeta)}\left.dg(u)/d\beta\right|_{\beta = \zeta} < 0 $. Reactive instability may potentially occur over regions where $ \sgn{(\zeta)}\left.dg(u)/d\beta\right|_{\beta = \zeta} = 0 $ such as over phases where $ \left.g(u)\right|_{\beta = \zeta} $ is a zero/nonzero constant, or over regions where $ \sgn{(\zeta)}\left.dg(u)/d\beta\right|_{\beta = \zeta} $ is nonzero but sufficiently small to be negligible in comparison to growth through reactive instability (if one exists). When $ \left.g(u)\right|_{\beta = \zeta} = 0 $, the interaction of the wave and the particles is non-resonant (as there are no particles with $ \beta = \zeta $ to resonate with the wave) and when $ \left.g(u)\right|_{\beta = \zeta} $ is a nonzero constant the interaction is resonant. \cite{M90_1990JPlPh..44..191M,M90_1990JPlPh..44..213M} considered a 3D J\"uttner distribution and concluded that temporal reactive growth is not possible when either the background or the beam interact resonantly with the wave; however, spatial reactive growth may be possible. For parameters relevant to pulsars, there is always resonant interaction between waves and particles in a J\"uttner distribution. Aside from an exceedingly small neighbourhood of $ \gamma_\phi \approx \gamma_{\rm b} $, Landau dissipation dominates any potential reactive growth and the dissipative term cannot be neglected.

The procedure we have employed in treating reactive instability analytically is based on the dispersion equation~\eqref{eq:K33r1}, and it seems that this cannot be justified, in the sense that one cannot justify neglecting the beam in determining the wave frequency. The contribution of the beam, which is heavily Landau damped, makes a significant contribution or even dominates the background invalidating the assumption $ |\omega_i/\omega| \ll 1 $. In effect, Landau damping dominates over reactive growth, invalidating the latter. We conclude that the conventional concept of a reactive instability does not apply for a J\"uttner distribution for parameter values relevant to pulsars.

\subsubsection{Interpretation based on Kramers-Kronig relations}

A formal interpretation for the absence of a reactive instability follows from the Kramers-Kronig relations, which imply that the real and imaginary parts of the response function ($K_{33}$ here) are Hilbert transforms of each other. In the conventional treatment of a reactive instability, one neglects the imaginary part of the response functions, implying (for consistency) that the real part is approximated by a form whose Hilbert transform is zero. Functions of the form $1/(\omega-\omega_0)$, where $\omega_0$ is independent of $\omega$, have Hilbert transform equal to zero, and the cold plasma dispersion equation (\ref{eq:cold}) can be written as a sum of functions of this form. However, our results suggest that, for the J\"uttner-distribution model for the beam, the imaginary part cannot be neglected in the region where the putative reactive instability is expected to occur. This suggests that the cold-plasma form of the dispersion equation cannot be a justifiable approximation, and that the dependence on $\omega$ in the numerator in (\ref{eq:K33r1_approx}) cannot be neglected. Our approximation assuming this numerator to be independent of $\omega$ is {\it a posteriori} not justifiable. When this dependence is included in our numerical results, we find no reactive instability, suggesting that the cold-plasma form is not a valid approximation and that the dispersion equation (\ref{eq:K33r1_approx}) cannot be approximated by a polynomial equation in $\omega$.

\section{Temporal and spatial growth in different frames}
\label{sect:LTgrowth}

The foregoing calculations of the growth rate are for temporal growth in the plasma rest frame ${\cal K}$. In applying the theory to a pulsar plasma, we are interested in spatial growth in the pulsar frame ${\cal K}'$. In this section we discuss the relations between temporal and spatial growth in different inertial frames. 

\subsection{Temporal and spatial growth}

The kinetic equation for growing waves is of the form of a total derivative, $D/Dt=\partial/\partial t +{\bi v}_{\rm g}\cdot{\bf\nabla}$ where ${\bi v}_{\rm g}$ is the group velocity, operating on the wave amplitude equal to $\Gamma$ times the wave amplitude, where $\Gamma$ is the growth rate calculated using dispersion theory. In the 1D case discussed here one has ${\bi v}_{\rm g}\cdot{\bf\nabla}=\beta_{\rm g}c\partial/\partial s$, where $\beta_{\rm g}$ is the group speed and $s$ is distance along the ray path. Temporal and spatial growth may be described by imaginary parts, $\Im\omega$ and $\Im k_\parallel$, of the frequency and (parallel) wavenumber, respectively, with the amplitude varying $\propto\exp(\Im\omega\,t-\Im k_\parallel\,s)$. The ratio of temporal and spatial terms depends partly on the boundary conditions: waves uniformly excited everywhere grow only in time and waves excited at a point source grow only in space away from the source. One has $\Im\omega=\Gamma, \Im k_\parallel=0$ for purely temporal growth and $\Im\omega=0, \Im k_\parallel=-\Gamma/c\beta_{\rm g}$ for purely spatial growth. 

A helpful step in relating temporal and spatial growth in different frames is to consider the frame in which the group speed is zero. In this frame the wave energy is not propagating, and wave growth must be purely temporal. In the plasma rest frame, the group velocity may be directed either outward or inward. Let us use labels $g\pm$ to denote these cases, writing the group velocity as $\pm c\beta_{\rm g}$, with $\beta_{\rm g}$ defined to be positive. There are then two frames ${\cal K}_{\rm g\pm}$, defined by the group velocity being zero for waves that are propagating outward and inward in ${\cal K}$, respectively. Let $t_{\rm g\pm}$ and $s_{\rm g\pm}$ be time and distance in ${\cal K}_{\rm g\pm}$. One has
\begin{equation}
t_{\rm g\pm}=\gamma_{\rm g}(t\pm\beta_{\rm g}s/c),
\qquad
s_{\rm g\pm}=\gamma_{\rm g}(s\pm\beta_{\rm g} ct),
\label{tgpm}
\end{equation}
where $\gamma_{\rm g}=(1-\beta_{\rm g}^2)^{-1/2}$ is the Lorentz factor corresponding to the group speed. The exponent in the variation of the amplitude, $\propto\exp(\Im\omega\,t-\Im k_\parallel\,s)$, is an invariant,\footnote{Using $ \Im k^\mu = [\Im\omega/c, 0, 0, \Im k_\parallel] $ and $ x^\mu = [ct, 0, 0, s] $ so that $ \Gamma = x^\mu \Im k_\mu = \Im\omega\, t - \Im k_\parallel s $.} and may be rewritten as a variation $\propto\exp(\Gamma t_{\rm g\pm}/\gamma_{\rm g})$. 

\subsection{Growth in the pulsar frame}

The growth rate in the pulsar frame ${\cal K}'$ is related to the growth rate in the rest frame of the plasma by a Lorentz transformation applied to $\Im\,\omega$, $\Im\,k_\parallel$, $t$ and $s$ in ${\cal K}$ to $\Im\,\omega'$, $\Im\,k'_\parallel$, $t'$ and $s'$ in ${\cal K}'$. The direct transforms are
\bea
\Im\,\omega'&=&\gamma_{\rm s}(\Im\,\omega+\beta_{\rm s}\Im\,k_\parallel c), 
\qquad
\Im\,k'_\parallel=\gamma_{\rm s}(\Im\,k_\parallel +\beta_{\rm s}\Im\,\omega/c),
\nn\\
\ms
t'&=&\gamma_{\rm s}(t+\beta_{\rm s}s/c),
\qquad\qquad\quad\;
s'=\gamma_{\rm s}(s+\beta_{\rm s}ct),
\label{LTIm}
\eea
and the inverse transforms follow by interchanging primed and unprimed quantities and replacing $\beta_{\rm s}$ by $-\beta_{\rm s}$. The group speeds $\pm\beta_{\rm g}$ in ${\cal K}$ transform into
\begin{equation}
\beta'_{\rm g\pm}=\frac{\pm\beta_{\rm g}+\beta_{\rm s}}{1\pm\beta_{\rm g}\beta_{\rm s}}
\label{LTbetag}
\end{equation}
in ${\cal K}'$. With the variation of the amplitude $\propto\exp(\Im\omega'\,t'-\Im k'_\parallel\,s')$ in ${\cal K}'$, one identifies $\Im\omega'=\gamma'_{\rm g\pm}\Gamma$, $\Im k'_\parallel=-\gamma'_{\rm g\pm}\Gamma/c\beta'_{\rm g\pm}$ with $\gamma'_{\rm g\pm}=\gamma_{\rm s}\gamma_{\rm g}(1\pm\beta_{\rm g}\beta_{\rm s})$, and with $\Gamma$ calculated in ${\cal K}$ by setting $\Im k_\parallel=0$ and $\Gamma=\Im\omega$.

\subsection{Growth factor}

The growth factor, $G$, such that the wave amplitude increases by a factor $\exp(G)$ during the wave growth, is an invariant. Suppose that (constant) wave growth starts at one event and finishes at another event, with the two events separated by $\Delta t,\Delta s$ in ${\cal K}$ and by $\Delta t',\Delta s'$ in ${\cal K}'$. Then one has
\begin{equation}
G=\Im\omega\Delta  t-\Im k_\parallel\Delta  s=\Im\omega'\Delta  t'-\Im k'_\parallel\Delta s',
\label{Gdef}
\end{equation}
which allows one to estimate the growth factor by applying appropriate boundary conditions in either frame.

One restriction on $G$ arises from the spatial variation of the number density $n'\propto1/r^3$ in the pulsar magnetosphere. This implies a characteristic distance $L_\parallel\approx3r/2$ over which the plasma frequency $\omega'_{\rm p} $ changes. (In the widely-favored ``sparking'' model, $L_\parallel$ is much smaller, of order the size of the cloud of pairs produced in a single spark.)  A wave growing in a bandwidth $\Delta\omega'_{\rm bw}$ in ${\cal K}'$ moves out of resonance, and ceases growing, after propagating a distance $\Delta s'=(\Delta\omega'_{\rm bw}/\omega')L_\parallel$. The foregoing results then enable one to estimate $G$. The estimate is quite different for waves propagating outward and inward in ${\cal K}$, both of which are propagating outward in ${\cal K}'$ for $\beta_{\rm s}>\beta_{\rm g}$. For $\gamma_{\rm s}\gg\gamma_{\rm g}$ one has $\gamma'_{\rm g+}\approx2\gamma_{\rm s}\gamma_{\rm g}$ and $\gamma'_{\rm g-}\approx\gamma_{\rm s}/2\gamma_{\rm g}$, implying that the spatial growth rate, $\gamma'_{\rm g\pm}\Gamma/c$, is of order $\gamma_{\rm s}^2$ larger for outward than for inward propagating waves in ${\cal K}$. The frequencies in ${\cal K}'$ are also quite different, $\omega'_+\approx2\gamma_{\rm s}\omega$ and $\omega'_-\approx\omega/2\gamma_{\rm s}\gamma_\phi^2$ for $\gamma_{\rm s}^2\gg\gamma_\phi^2$. As a result the fractional growth rate in ${\cal K}'$ is approximately equal to the fractional growth rate in ${\cal K}$ for outward propagating waves, but for inward propagating waves it is larger by a factor of order $\gamma_{\rm s}^2\gamma_\phi^2/\gamma_{\rm g}^2$. A detailed comparison of these two cases is needed, and we plan to discuss these details elsewhere. Suffice it to say here, neither seems favorable as a basis for pulsar radio emission.

\section{Discussion}
\label{sect:discussion}

Our objective in this series of papers is to present a critical discussion of beam-driven instabilities in a pulsar plasma, that is, in a highly relativistic, strongly magnetized, 1D pair plasma. We assume that all particle distribution functions are either non-streaming or streaming J\"uttner distributions, with $ \rho_0 \lesssim 1 $. We discuss wave dispersion in such a plasma in the plasma rest frame ${\cal K}$ in Paper~1 and discuss the transformation to the pulsar frame ${\cal K}'$ in Paper~2. In this paper we discuss growth rates for various possible beam-driven instabilities. We find that several difficulties, discussed in Papers~1 and~2, with beam-driven wave growth are further compounded by a quantitative difficulty: the growth rate is too small to be effective.  This leads us to conclude that the widely-held assumption that a beam-driven instability plays a central role in the generation of pulsar radio emission is not justified for parameters considered plausible. Before commenting on this conclusion, we summarize the difficulties identified.  

\subsection{Difficulties identified in Papers~1 and~2}
It is widely assumed that pulsar radio emission involves Langmuir-like waves growing through a beam-driven instability. As discussed in Paper 1, in RPE these waves are assumed to produce escaping radiation through nonlinear (or other) processes, and in CCE these waves are assumed to develop into solitons, which provide the particle bunches postulated for curvature emission to be coherent. A major difficulty, pointed out in Paper~1, is that there are no Langmuir-like waves in pulsar plasma. For the particular case of parallel propagation the dispersion relations reduce to $\zeta=\beta_A$ for  the Alfv\'en and X~modes and to $\omega=\omega_{L0}(\zeta)$ for the L~mode.\footnote{We identify the distribution discussed in Paper~1 with the background distribution.} Although L~mode waves exist for all superluminal phase speeds, $\zeta>1$, only subluminal waves are relevant for a beam instability. Subluminal L~mode waves exist for only a tiny range of phase speeds, $\zeta\approx1-1/2\gamma_\phi^2$ with $\gamma_\phi\gg1$. Specifically, there is no solution for $\gamma_\phi\lesssim2\langle\gamma\rangle_0$, and for  $2\langle\gamma\rangle_0 \lesssim \gamma_\phi \lesssim \gamma_{\rm m} \approx 6\langle\gamma\rangle_0 $, the waves have anomalous dispersion (e.g., implying negative energy and other unusual features). The only relevant (parallel-propagating) waves are those in the range $\gamma_{\rm m} < \gamma_\phi < \infty $. Over this tiny ($\approx0.13/\langle\gamma\rangle_0^2$) range of $\zeta$, $\omega_{L0}(\zeta)$ changes significantly implying, inter alia, a large (negative) value of ${\rm d}\omega_{L0}(\zeta)/{\rm d}\zeta$, leading to a very small ratio of electric to total energy, $R_L(\zeta)\approx1/24\langle\gamma\rangle_0^2$. The resulting wave properties (for $\langle\gamma\rangle_0 \gg 1$) are inconsistent with any plausible definition of ``Langmuir-like''.  An exception is that the L~mode has longitudinal polarization, but only for strictly parallel propagation: for slightly oblique propagation, the parallel L~mode becomes the O~mode for $\zeta>\beta_A$ ($\gamma_\phi>\gamma_A$) and the Alfv\'en mode for $\zeta<\beta_A$ ($\gamma_\phi<\gamma_A$).

A second difficulty, discussed in Papers~1 and~2, concerns how the beam is formed. The favored suggestion \citep{U87,UU88} is based on a ``sparking'' model in which the pair creation is structured in both space and time, resulting in ``clouds'' of pairs. A beam instability is attributed to the counter-streaming motion resulting when faster particles in a following cloud over take slower particles in a preceding cloud \citep{U87,AM98}. In Paper~2 it is pointed out that this model requires that the two distributions (e.g. beam and background) be well separated, in the sense that the combined distribution function, $g(u)$, has a well-defined minimum between them, so that there is a positive gradient ${\rm d}g(u)/{\rm d}u>0$ above the minimum. For a beam propagating through a background plasma, which is the case usually assumed, this separation condition requires $\gamma_{\rm b} \gtrsim \gamma_{\rm b, min} \sim 10\av{\gamma}_0$. This condition is not plausibly satisfied for a ``sparking'' model, and some more extreme assumption, than random differences between clouds, is required to account for counter-streaming. 

A further complication relates to the resonance condition $\gamma_\phi=\gamma_{\rm b}$. For $\gamma_{\rm b}>\gamma_A$, the resonance is in the branch that becomes the O~mode for oblique propagation, and for $\gamma_{\rm b}<\gamma_A$  the resonance is in the branch that becomes the Alfv\'en mode for oblique propagation. The very large value of $\gamma_A$ in a pulsar plasma suggests that any beam-driven instability is in the Alfv\'en mode, as earlier authors suggested \citep{TK72,Letal82,L00}. However, the conventional dispersion relation for the Alfv\'en mode, $z=\beta_A$ or $\gamma_\phi=\gamma_A$ in the notation used here, is relevant only for $\gamma_{\rm b}\approx\gamma_A$, and the dispersion relation corresponding to $\gamma_\phi=\gamma_{\rm b}<\gamma_A$ is L-mode-like, as discussed in Paper~1.

\subsection{Choice of beam distribution}

An assumption that is widely made is that the beam is described by a relativistically streaming Gaussian distribution, $g(u)\propto\exp[-(u-u_{\rm b})^2/{u_{\rm Tb}^2}]$, cf.\ (\ref{Gaussian}), with $u_{\rm Tb}\approx\langle\gamma\rangle\gg1$. We argue that such a distribution is artificial, and that a realistic relativistic streaming distribution is obtained by Lorentz transforming a (e.g., J\"uttner, Gaussian, water-bag or bell) distribution in the rest frame to the (primed) frame in which the particles are streaming at $\beta_{\rm b}$. The distribution function transforms from $g(u)$ to $g'(u')=g(u)$, with $u'=\gamma_{\rm b}\gamma(\beta-\beta_{\rm b})$, which is very much broader that the assumed relativistically streaming distribution. For example, \citet{AM98} assumed a relativistically streaming distribution and expressed it in dimensionless variables, $p_d=\gamma/\gamma_{\rm b}$, $p_{dT}=u_{\rm Tb}/u_{\rm b}$ for $u_{\rm b},u_{\rm Tb}\approx\gamma,\gamma_{\rm Tb}\gg1$, giving $g(u)\propto\exp[-(p_d-1)^2/p_{dT}^2$. This choice may be compared with the distributions obtained by assuming a J\"uttner distribution, $\propto\exp[-\gamma/\langle\gamma\rangle]$, or a relativistic Gaussian distribution, $\propto\exp[-u^2/{u_{\rm Tb}^2}]$, and Lorentz transforming to include the streaming motion giving, in dimensionless variables, $g(u)\propto\exp[-(p_d-1)^2/2p_d\langle\gamma\rangle\gamma_{\rm b}]$ and $g(u)\propto\exp[-\{(p_d-1)/p_{dT}\}^2\{(p_d+1)/2p_d\gamma_{\rm b}\}^2]$, respectively. The additional factors $\gamma_{\rm b}$ and $\gamma_{\rm b}^2$, respectively, in the denominators in the exponents imply that the Lorentz-transformed distribution are of order $\gamma_{\rm b}$ broader than these distributions in their rest frame.  We regard the assumption of a relativistically streaming distribution, rather than a Lorentz-transformed distribution, as artificially narrow, thereby obscuring the difficulty of satisfying the separation condition. 
\subsubsection{Kinetic instability}

In this paper, we ignore these difficulties and estimate the magnitude of the growth rate, which we express as a fraction of the wave frequency. The growth rate $ \overline{\Gamma}_L(\zeta) $ corresponds to the parallel L~mode which is larger than the growth rate for oblique modes (e.g. O and Alfv\'en modes). As discussed by ELM, the instability can take two different forms, kinetic or reactive, with the former applying only if the growth rate is less than the bandwidth of the growing waves. For a relativistically streaming distribution ELM found that this condition is not satisfied and that any instability must be reactive. With our choice of distribution function, specifically one constructed by including the streaming through a Lorentz transformation, the much larger width of the distribution implies that the (maximum) growth rate is reduced and that the bandwidth of the growing waves is increased. As a result we find that this condition for the kinetic version to apply can be marginally satisfied.  However, the relatively small growth rate introduces another difficulty: an essential condition for effective wave growth is a large growth factor $G$, and our estimates suggest $G<1$. A beam-driven instability is then ineffective in generating any waves in pulsar plasma.

\subsubsection{Reactive instability}

In \S\ref{sect:reactive} we outline a conventional procedure for reducing the dispersion equation for a warm beam to a cold-plasma-like form that is a quartic equation in $\omega$. This form is the basis for the usual treatment of a reactive beam-driven instability. We suggest seemingly plausible assumptions to reduce the dispersion equation for a J\"uttner model for the beam to a quartic equation, and write down approximate solutions for growing waves. However, our numerical calculation fails to support this procedure, and we find no reactive instability. Our interpretation is that the imaginary part of the contribution of the beam to the dispersion equation cannot be neglected, and that this invalidates the approximations made in reducing this equation to a cold-plasma-like form.

\section{Conclusions}
\label{sect:conclusions}

Our main conclusion is that beam-driven wave growth does not occur for plausible parameters in a pulsar plasma. The kinetic instability can occur, but the growth rate is too small to be effective, and the reactive instability does not occur. This implies either that RPE and CCE are untenable as pulsar radio emission mechanisms, or that the model on which our ``plausible'' parameters are based is incorrect in some important way. We note that the assumption that has the greatest negative effect on possible beam-driven wave growth is that the particles have a relativistic spread in their rest frame, $\langle\gamma\rangle_0 -1\gtrsim 1$. For example, if this assumption is replaced by $\langle\gamma\rangle_0 - 1 \ll 1$, the L~mode in the rest frame becomes Langmuir-like, as assumed by \citet{W94} in the model used by \citet{EH16} in arguing that RPE can account for the nanoshots observed in the Crab pulsar. However, such a nonrelativistic spread is inconsistent with models for pair cascades. 

The implications for pulsar radio emission need to be discussed critically in the context of a specific model for a pulsar magnetosphere. We propose to discuss the implications for pulsar radio emission in detail elsewhere.

\section*{Acknowledgments} 
We thank Mike Wheatland for comments on the manuscript.
The research reported in this paper was supported by the Australian Research Council through grant DP160102932.

\bibliographystyle{jpp}

\bibliography{Pulsar_radio_Refs}

\appendix

\section{Penrose criterion for instability: derivation}
\label{app:Penrose}

For a wave to grow exponentially we must have $\Im\omega > 0 $. Hence for instability to be possible we require that at least one solution of $ K_{33}(\omega, k_\parallel) = 0 $ lies in the upper half of the complex $\omega$-plane. Consider the function~\citep{M86}
\begin{equation}
    G(\omega, k_\parallel) = \frac{\partial K_{33}(\omega, k_\parallel)/\partial \omega}{K_{33}(\omega, k_\parallel)},
\end{equation}
so that solutions of $ K_{33}(\omega, k_\parallel) = 0 $ correspond to poles of $G(\omega, k_\parallel)$. By contour integration, performed in a positive (counter-clockwise) sense,
\begin{equation}
    \oint_{c_\omega}d\omega\, G(\omega, k_\parallel) =
    \oint_{c_\omega}d\omega\, \frac{\partial K_{33}(\omega, k_\parallel)/\partial \omega}{K_{33}(\omega, k_\parallel)}
    = \oint_{c_K} \frac{dK_{33}}{K_{33}}
    = 2\pi i N,
\end{equation}
where the contour $ c_\omega $ encloses the entire upper half of the complex $\omega$-plane, contour $ c_K $ is the map of $ c_\omega $ in the complex $K_{33}$-plane, and $ N $ is an integer equal to the number of zeros of $ K_{33}(\omega, k_\parallel) $ contained by the contours. We comment that for $ N $ to be nonzero the contour $ c_K $ must encircle the origin in the complex $ K_{33}$-plane. Our aim is to derive a necessary and sufficient condition for $ c_K $ to encircle the origin and thus ensure that there is a growing solution ($ \Im\omega > 0 $) of $ K_{33}(\omega, k_\parallel) = 0 $.

We may write $ K_{33}(\omega, k_\parallel)$ as
\begin{equation}\label{eq:K33exp}
    K_{33}(\omega, k_\parallel)
    = 1 - \frac{\omega_{\rm p0}^2}{c^2k_\parallel^2}W(z)
    = 1 - \frac{\omega_{\rm p0}^2}{c^2k_\parallel^2}\frac{1}{n_0}\int_{-1}^{1}d\beta\frac{dg(u)/d\beta}{\beta-\omega/ck_\parallel},
\end{equation}
where $ z = z(\omega, k_\parallel) = \omega/ck_\parallel $. For simplicity in discussion, we assume $ 0 < k_\parallel < \infty $ in the following so that $ \sgn{(z)} = \sgn{(\omega)} $. We may write the above, for $ -ck_\parallel < \omega < ck_\parallel $ (or $ -1 < z < 1 $), as
\begin{equation}\label{eq:K33wk}
\begin{split}
    K_{33}(\omega, k_\parallel) 
    = 1 &+ \frac{\omega_{\rm p0}^2}{c^2k_\parallel^2}\frac{1}{n_0}\Bigg[\left.2\gamma^2 g(u)\right|_{\beta = z}
    + \wp\int_{-1}^{1}d\beta\frac{\left.g(u)\right|_{\beta = z} - g(u)}{(\beta-z)^2}\Bigg]\\
    &+ i\frac{\pi\omega_{\rm p0}^2 \rho_0 \gamma_\phi^3}{2c^2k_\parallel^2 K_1(\rho_0)}\frac{1}{n_0}\left[ze^{-\rho_0\gamma_\phi} + \varepsilon_n\varepsilon_\rho\varepsilon_K\gamma_{\rm b}(z-\beta_{\rm b})e^{-\rho_b\gamma_{\rm b}\gamma_\phi(1-z\beta_{\rm b})}\right].
\end{split}
\end{equation}
The semi-circle section of the contour $ c_\omega $ at infinity is mapped to unity on the contour $ c_K $ as evident from~\eqref{eq:K33exp}. Now, consider the portion of $ c_\omega $ along the real axis of the complex $ \omega$-plane. As $ \omega \to -ck_\parallel $ (or $ z \to -1 $) from below, we have
\begin{equation}
    \lim_{\omega \to -ck_\parallel}\Re K_{33}(\omega, k_\parallel) 
    = 1 - \frac{\omega_{\rm p0}^2}{c^2k_\parallel^2} \Re W(-1) 
    \approx 1 - \frac{\omega_{\rm p0}^2}{c^2k_\parallel^2}\langle\gamma\rangle_0\left[2 + \frac{\varepsilon_n}{2\varepsilon_\rho\gamma_{\rm b}^2}\right] < 1,
\end{equation}
from above and $ \Im K_{33}(\omega, k_\parallel) $ departs from $ -i0 $ at $ \omega = -ck_\parallel $. As $ \omega \to ck_\parallel $ (or $ z \to 1 $) from below, we obtain 
\begin{equation}
    \lim_{\omega \to +ck_\parallel}\Re K_{33}(\omega, k_\parallel) 
    = 1 - \frac{\omega_{\rm p0}^2}{c^2k_\parallel^2} \Re W(1) 
    \approx 1 - \frac{\omega_{\rm p0}^2}{c^2k_\parallel^2}\langle\gamma\rangle_0\left[2 + 8\frac{\varepsilon_n \gamma_{\rm b}^2}{\varepsilon_\rho}\right] < 1,
\end{equation}
and $ \Im K_{33}(\omega, k_\parallel) $ approaches $ +i0 $ at $ \omega = ck_\parallel $ and remains zero for $ \omega > ck_\parallel $ (or $z > 1$). As $ \omega \to \infty $ we have $ K_{33}(\omega, k_\parallel) \to 1 $ closing the contour. From~\eqref{eq:K33exp} and~\eqref{eq:K33wk} we see that $ \Im K_{33}(\omega, k_\parallel) = 0 $ when $ dg(u)/d\beta = 0 $ with the turning points confined to $ 0 < \omega < ck_\parallel\beta_{\rm b} $ (or $ 0 < z < \beta_{\rm b} $); and $ \Im K_{33}(\omega, k_\parallel) $ switches sign across the turning points of $ g(u) $. 

Suppose that the distributions are well-separated with turning points at $ z_1$, $z_2 $ and $z_3$ and with $ z_1 < z_2 < z_3 $ (or equivalently $ \omega_1 < \omega_2 < \omega_3 $). Then we must have $ \left.g(u)\right|_{\beta = z_2}$ as the global minimum (ignoring $ \beta = \pm 1 $) and $ g(z_1)$ and $\left.g(u)\right|_{\beta = z_3}$ as local/global maxima. These imply that the contour $ c_K $ starts at $ 1 - i0 $ crosses the real axis, in order, at $ z_1 $ (upwards), $ z_2 $ (downwards) and $ z_3 $ (upwards) and then returns to $ 1 + i0 $; the crossing is downwards at the minimum of $ g(u) $ and upwards at its maxima.

For sufficiently large $ u_{\rm b}>0 $, such that the distribution function $ g(u) $ has a minimum between the two peaks, we may write $ z_1 \approx \delta_1 $ and $ z_3 \approx \beta_{\rm b} - \delta_2 $ with $ 0 < \{\delta_1, \delta_2\} \ll 1 $. This allows us to make the approximations
\begin{align}
    W(z_1)
        & \approx \Re W_0(0) + \frac{\varepsilon_n}{4\gamma_{\rm b}^2}\Re W''_1(-\beta_{\rm b})
        \approx \frac{\varepsilon_n}{4\gamma_{\rm b}^2}\Re W''_1(\beta_{\rm b}) - \frac{1}{\langle\gamma\rangle_0},\\
    W(z_3) 
        & \approx \Re W_0(\beta_{\rm b}) + \varepsilon_n\gamma_{\rm b} \Re W''_1(0)
        \approx \Re W_0(\beta_{\rm b}) - \frac{\varepsilon_n\varepsilon_\rho}{\gamma_{\rm b}^2\langle\gamma\rangle_0},
\end{align}
where we use $ \Re W_\alpha(0) = -\rho_\alpha \approx -1/\langle\gamma\rangle_\alpha $ for $ \rho_\alpha \ll 1$. For $ \gamma_\phi \lesssim 2 \langle\gamma\rangle_\alpha $ we have $ \Re W_\alpha(z) \lesssim 0 $ with a minimum value of approximately $ -0.36 \langle\gamma\rangle_\alpha $ at $ \gamma_\phi \approx \langle\gamma\rangle_\alpha $. The maximum value of $ \Re W_\alpha(z) \approx 2.7 \langle\gamma\rangle_\alpha $ occurs at $ \gamma_\phi \approx 6 \langle\gamma\rangle_\alpha $ which reduces to $ \approx 2\langle\gamma\rangle_\alpha $ at $ z = 1$. We then have $ \Re W(z_1) < 0 $ for $ \gamma_{\rm b} \gtrsim (\varepsilon_n/\varepsilon_\rho)^{1/2}\av{\gamma}_0 $, which is well satisfied for separated distributions with $ \gamma_{\rm b} > \gamma_{\rm b, min} \sim 10\av{\gamma}_0 $. This implies that $ \Re K_{33}(\omega_1, k_\parallel) > 0 $ for relevant parameters. At $ z_3 $ we have $ \Re W(z_3) < 0 $ if $ \gamma_{\rm b} \lesssim (\varepsilon_n\varepsilon_\rho)^{1/2}/2\av{\gamma}_0 $ which is never satisfied for well separated distributions. From~\eqref{eq:K33wk} it is evident that $ \Re W(z_2) > \Re W(z_3) $ since $ \left.g(u)\right|_{\beta = z_3} > \left.g(u)\right|_{\beta = z_2} $ and $ \left.\gamma\right|_{\beta = z_3} > \left.\gamma\right|_{\beta = z_2} $ as $ z_3 > z_2 $. Thus we have $ \Re K_{33}(\omega_3, k_\parallel) > \Re K_{33}(\omega_2, k_\parallel) $. 

Therefore, there is a small $ k_\parallel = k_0 $ with $ \Re K_{33}(\omega_2, k_0) < 0 $. If $ \Re K_{33}(\omega_3, k_0) > 0 $ then we may adjust $ k_0 $ such that $ \Re K_{33}(\omega_3, k_0) > 0 $ while $ \Re K_{33}(\omega_2, k_\parallel) < 0 $ since $ \Re W(z_2) > \Re W(z_3) $. The contour $ c_K $ may then cross the negative real axis in the complex $ K_{33} $-plane only once, at $ \Re K_{33}(\omega_2, k_0) = 1 - (\omega_{\rm p0}^2/c^2k_0^2) \Re W(z_2) $, and must encircle the origin. Therefore, a necessary and sufficient condition for existence of an instability is $ \Re W(z_2) > 0 $ or
\begin{equation}
    \int_{-1}^{1}d\beta\frac{dg(u)/d\beta}{\beta - z_2 } = -\left.2\gamma^2g(u)\right|_{\beta = z_2} - \wp\int_{-1}^{1}{\rm d}\beta\, \frac{\left.g(u)\right|_{\beta = z_2} - g(u)}{(\beta - z_2)^2} > 0,
\end{equation}
where $ g(u) $ is a global minimum at $ \beta = z_2 $. This is the Penrose criterion for instability: an instability exists if $ \Re W(z_2) > 0 $ when $ \Im W(z) $ switches sign at $ z = z_2 $. When plotting $ \Re W(z) $ against $ \Im W(z) $ the plot starts at $ + i0 $ crossing the real axis, in order, at $ z_1 $ (downwards), $ z_2 $ (upwards) and $ z_3 $ (downwards) and then returning to $ - i0 $.

\begin{figure}
\centering
\psfragfig[width=1.0\columnwidth]{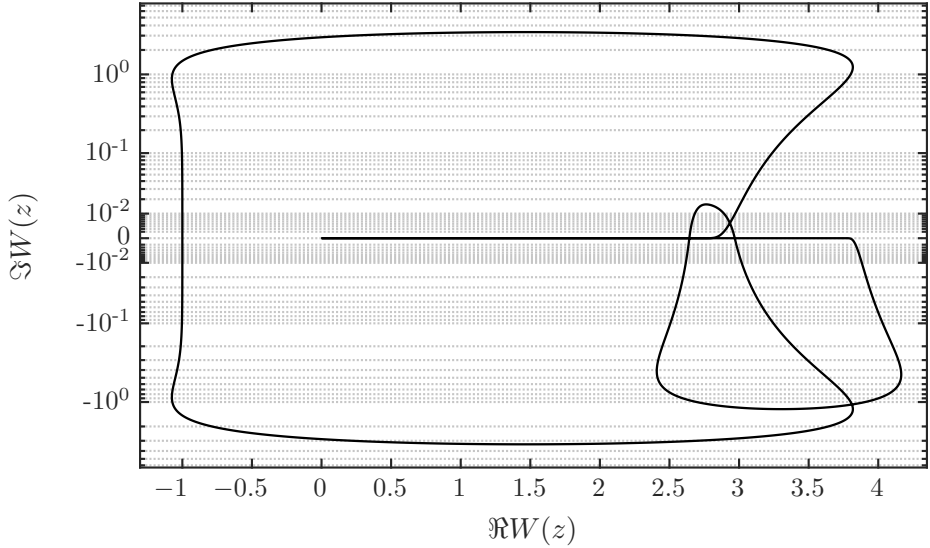}
 \caption{Plots of $ \Im W(\omega, k_\parallel) $ against $ \Re W(\omega, k_\parallel) $ for $ \rho = 1 $, $ \varepsilon_\rho = 1 $, $ \varepsilon_n = 10^{-4} $, $ \gamma_{\rm b}/\gamma_{\rm b, min} = 2 $. The y-axis has been scaled using a bi-symmetric log transformation $ x \mapsto \sgn{(x)}\log_{10}\left(1 + |x/10^C|\right) $ with $ C = -2 $ so that the range $ 0 $ to 1 is presented by two decades.}
 \label{fig:Penrose1}  
\end{figure}

Figure~\ref{fig:Penrose1} shows a plot of $ \Im W(\omega, k_\parallel) $ against $ \Re W(\omega, k_\parallel) $ for $ \rho = 1 $, $ \varepsilon_\rho = 1 $, $ \varepsilon_n = 10^{-4} $, $ \gamma_{\rm b}/\gamma_{\rm b, min} = 2 $. The y-axis has been scaled using a bi-symmetric log transformation $ x \mapsto \sgn{(x)}\log_{10}\left(1 + |x/10^C|\right) $ with $ C = -2 $ so that the range $ 0 $ to 1 is presented by two decades. This curve may be considered as a map of the contour $ c_\omega $ (or $ c_K $) onto the complex $ W $-plane: contour $ c_W $. The curve starts at $ (\Re W(z), \Im W(z) ) = (0, 0) $ when $ \omega = -\infty $, moves to the right along the real axis as $ \omega \to -ck_\parallel $ (or $ z = -1 $) where $ \Im W(z) $ has a nonzero positive value [around $ (\Im W(z), \Re W(z)) \approx (+0i, 2.8)$]. The curve goes up moving anti-clockwise and crossing the real axis at $ z = z_1 \approx 0 $ (downwards) with $ \Re W(z_1) \approx -1 < 0 $. Continuing along, the curve eventually crosses the real axis at $ z = z_2 $ (upwards), with $ \Re W(z_2) \approx 3 > 0 $, and at $ z = z_3 \approx \beta_{\rm b} $ (downwards), with $ \Re W(z_3) \approx 2.6 < \Re W(z_2) $, looping back towards the real axis at $ \omega = -ck_\parallel $ (or $ z = 1 $) [around $ (\Im W(z), \Re W(z)) \approx (-0i, 3.8)$] and approaching the origin as $ \omega \to \infty $. The mode corresponding to these parameter values is then unstable by the Penrose criterion.

\end{document}